\documentclass[preprint2]{proto}
\usepackage{times}
\usepackage{psfig}

\newcommand{\refs}{\par\noindent\hangindent=1pc\hangafter=1}

\newcommand{\ltappeq}{\raisebox{-0.6ex}{$\,\stackrel
{\raisebox{-.2ex}{$\textstyle <$}}{\sim}\,$}}

\newcommand{\mdot}{\mbox{$\stackrel{.}{M}$}}

\begin{document}

\title{\textbf{\LARGE Ultra-Compact H~II Regions and the Early Lives of Massive Stars}}

\author {\textbf{\large M. G. Hoare}}
\affil{\small\em University of Leeds}

\author {\textbf{\large S. E. Kurtz and S. Lizano}}
\affil{\small\em Universidad Nacional Aut\'{o}noma de M\'{e}xico - Morelia}

\author {\textbf{\large E. Keto}}
\affil{\small\em Harvard University}

\author {\textbf{\large P. Hofner}}
\affil{\small\em New Mexico Institute of Technology and National Radio Astronomy Observatory}

\begin{abstract}
%\begin{list}{ } {\rightmargin 1in}
%{\leftmargin 0in}
\baselineskip = 11pt
\leftskip = 0.65in
\rightskip = 0.65in
%rule{4.75in}{0.5pt}
%\vskip 1pt
\parindent=1pc {\small We review the phenomenon of ultra-compact H~II
regions (UCHIIs) as a key phase in the early lives of massive
stars. This most visible manifestation of massive star formation
begins when the Lyman continuum output from the massive young stellar
object becomes sufficient to ionize the surroundings from which it was
born. Knowledge of this environment is gained through an understanding
of the morphologies of UCHII regions and we examine the latest
developments in deep radio and mid-IR imaging. SPITZER data from the
GLIMPSE survey are an important new resource in which PAH emission and
the ionizing stars can be seen.  These data provide good indications
as to whether extended radio continuum emission around UCHII regions
is part of the same structure or due to separate sources in close
proximity. We review the role played by strong stellar winds from the
central stars in sweeping out central cavities and causing
the limb-brightened appearance. New clues to the wind properties from
stellar spectroscopy and hard X-ray emission are discussed.  A range
of evidence from velocity structure, proper motions, the molecular
environment and recent hydrodynamical modeling indicates that cometary
UCHII regions require a combination of champagne flow and bow shock
motion.  The frequent appearance of hot cores, maser activity and
massive young stellar objects (YSOs) ahead of cometary regions is
noted. Finally, we discuss the class of hyper-compact H~II regions or
broad recombination line objects. They are likely to mark the
transition soon after the breakout of the Lyman continuum radiation
from the young star. Models for these objects are presented, including
photo-evaporating disks and ionized accretion flows that are
gravitationally trapped.  Evolutionary scenarios tracing young massive
stars passage through these ionized phases are discussed.
\\~\\~\\~}%leave this in to get the correct vertical space after the abstract

%\end{list}
\end{abstract}  

\section{\textbf{INTRODUCTION}}

The time when newly formed massive stars begin to ionize their
surroundings is one of the energetic events that underlines their
important role in astrophysics. As they evolve, the copious amounts of
UV radiation and powerful stellar winds they produce have a profound
effect on the surrounding interstellar medium.  Their early lives are
spent deeply embedded within dense molecular cores whose high column
densities absorb the optical and near-IR light from the young stars,
shielding them from view.  One of the first observable manifestations
of a newly formed massive star is the radio free-free emission of the
H~II region surrounding the star. Since only the most massive stars
produce significant radiation beyond the Lyman limit, embedded H~II
regions are a unique identifier of high mass star formation.

The absorption of the UV and optical radiation by dust, both in and
outside the newly formed nebula, heats the grains to temperatures
that range from the sublimation temperature close to the star to
interstellar temperatures in the surrounding molecular cloud.  Owing
to the high luminosity of the massive stars, H~II regions are some of
the strongest infrared sources in the galaxy.  Similar to the radio
emission, the thermal IR radiation is little affected by extinction.
Thus the combination of the radio and IR wave bands allows us to peer
deep into the star forming clouds to study the processes of star
formation within.
 
The youngest massive stars are associated with the smallest H~II
regions. These are the ultra-compact H~II regions (UCHII) and the newly
identified class of hyper-compact (HCHII) regions.  UCHII regions were
first distinguished from merely ``compact'' H~II regions around 25
years ago, and came to be defined observationally ({\em Wood and
Churchwell}, 1989a) as those regions with sizes $\leq 0.1$~pc,
densities $\geq 10^4$~cm$^{-3}$, and emission measures $\geq
10^7$~pc~cm$^{-6}$. Since then, hundreds of UCHII regions with these
general properties have been found. Whilst the division of H~II
regions into different degrees of compactness may be a convenient
label, for the larger objects at least, it is likely to have little
physical significance.  Once in the expansion phase, the
physics of their dynamics probably stays the same until the molecular
material is cleared away and the OB star joins the field population.
Of greater interest are the smallest H~II regions, as they 
more to tell us about the process of massive star formation.  The
ionized gas within the UCHII and HCHII regions not only reveals
properties of the stars themselves, but also lights up the immediate
surroundings to allow investigations of the density distribution and
environment into which the massive stars are born. The external
environment has a profound influence on the evolution of the H~II
regions.

The properties of UCHII regions and their immediate precursors have
been reviewed previously by {\em Churchwell} (2002) and {\em Kurtz et
al.} (2000). In this review we concentrate on more recent developments
in the field of UCHII and HCHII regions.  These include new theories
of massive star forming accretion flows and the new views of UCHII
regions opened up by infrared studies on large ground-based telescopes
and the SPITZER satellite. The SPITZER GLIMPSE survey has covered a
large part of the inner galactic plane at unprecedented spatial
resolution and depth in the 4-8~$\mu$m region where there is a local
minimum in the extinction curve.  High resolution X-ray studies with
{\em Chandra} are also beginning to bear on the problem. Together
these promise great new insights into how OB stars are formed and
interact with their environment.

\bigskip

\centerline{\textbf{ 2. ULTRA-COMPACT H~II REGIONS}}
\bigskip

\noindent
\textbf{2.1 Morphologies}
\bigskip

%\noindent
%\textbf{2.1.1 Compact Emission}
%\bigskip

The morphologies of UCHII regions are important since they yield clues
to the state of the surrounding medium relatively soon after a massive
star has formed. The common appearance of a regular morphology
indicates that there are ordered physical processes occurring
during massive star formation rather than just stochastic ones.

\begin{figure*}
 \epsscale{2.0}
\plotone{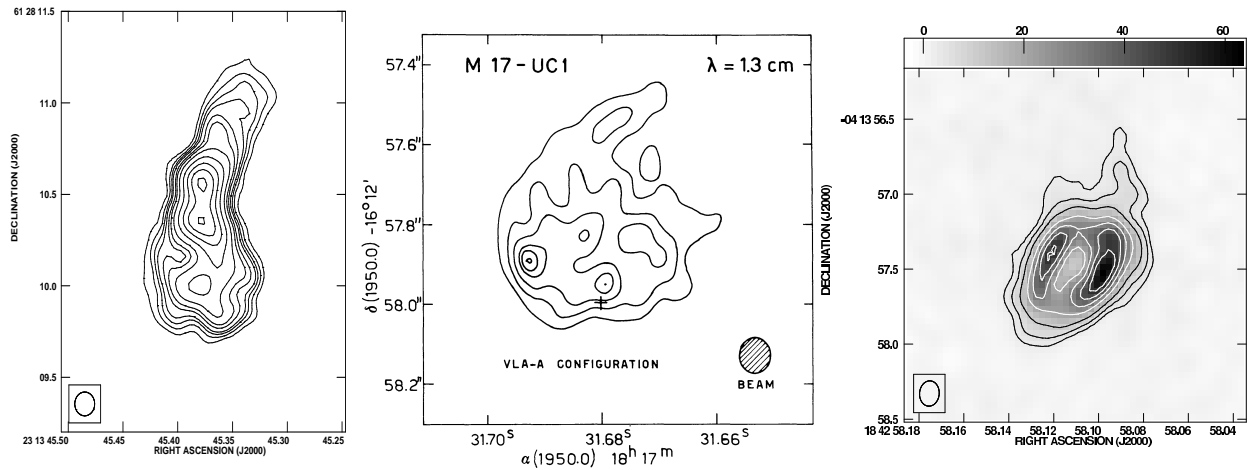}
  \caption{The different morphologies of hyper-compact H~II regions. Left:
the bipolar NGC 7538 IRS 1 ({\em Franco-Hern\'{a}ndez and Rodr\'{i}guez}, 2004), Middle: the cometary M17 UC1 ({\em Felli et al.}, 1984) and Right: the shell-like G28.20$-$0.04 ({\em Sewilo et al.}, 2005).}
\label{fig:HCHII}
\end{figure*}

As part of their pioneering high resolution radio surveys of massive
star forming regions {\em Wood and Churchwell} (1989a) developed a morphological classification scheme for
UCHII regions. Together with {\em Kurtz
et al.} (1994), they found that 28\%
of UCHII regions are spherical, 26\% cometary, 26\% irregular, 17\%
core-halo and 3\% shell. The significant numbers of unresolved sources
have been omitted here, since nothing can be said about their
morphology. {\em Walsh et al.} (1998) studied a sample of southern UCHII
regions and classified most of their sources as either cometary (43\%) or
irregular (40\%) after omitting the unresolved ones. As pointed
out by {\em Wood and Churchwell}, many of the sources classified as
spherical do reveal ordered morphology when observed at higher spatial
resolution. Examples are
M17-UC1, which was shown to be cometary ({\em Felli et al.}, 1984) and
G28.20-0.04, which is shell-like ({\em Sewilo et al.}, 2004) (see Fig.
\ref{fig:HCHII}). If the spherical sources of earlier studies were
also discounted, then the proportion of cometaries in the well-resolved
objects would be similar to that from {\em Walsh et al.}

As {\em Wood and Churchwell} forewarned, radio
interferometric observations have a limited range of spatial
scales that they are sensitive to. The larger objects can quickly
become over-resolved and break up into irregular sources.  The
snapshot nature of observations necessary to investigate significant
numbers of sources also limits the dynamic range. For cometary objects
this often means that at high resolution only the dense material at
the head is seen and the full extent of the much weaker emission in
the tail is not.  Even for relatively simple objects like the
archetypal cometary UCHII region, G29.96-0.02, the complete radio
continuum picture is only fully revealed by time consuming deep,
multi-configuration observations ({\em Fey et al.}, 1995).

The recent radio studies by {\em De Pree et al.} (2005) also address
this point.  They have conducted deep, multi-configuration studies of
two regions of intense massive star formation, W49A and Sgr B2, where
nearly 100 UCHII regions are found.  These radio data had good spatial
and dynamic range leading {\em De Pree et al.} to reassess the
morphological classes. They classify about one third of their sources
as 'shell-like', after omitting the unresolved fraction. At least half
of these are very asymmetric and could just as easily be classified as
cometary. This would take the total cometary fraction to over a third
in that sample.  The high dynamic range of the {\em De Pree et al.}
data also led them to drop the core-halo class and instead attempt to
ascertain the shape of the compact and extended emission
separately. These sources often appear to be a superposition of a
compact source on a different, more extended source.  They also detect
a significant population of UCHII regions elongated along one
axis. Following {\em Churchwell} (2002) they classify these regions as
bipolar. It is not yet clear whether the detection of these bipolar
sources is due to the better quality radio imaging or the more extreme
pressure in these two environments compared to the general galactic
population.

Imaging in the infrared is sensitive to all spatial scales at once, from
the resolution limit up to the total size of the image. This could be
several orders of magnitude of spatial dynamic range as long as the
image is sensitive enough to pick up very extended, low surface
brightness emission. Therefore, IR imaging can in principle overcome some
of the limitations of radio interferometric snapshot data for
morphological classification, although it also comes with its own
disadvantages. In the near-IR, the continuum emission from UCHII
regions is mostly made up of bound-free and free-free emission from
the nebular gas. There are also minor contributions from scattered light, emission
from very hot grains and non-thermal equilibrium emission from very
small grains. Hence, near-IR images should show the same morphology as
the radio, apart from the effect of intervening extinction.  An
example of this is seen in the near-IR image of G29.96-0.02 by {\em
Fey et al.}, where the nebular continuum looks just like their deep
multi-configuration radio map.  However, in general the extinction in
the near-IR is often too high and renders the UCHIIs invisible or cut
through by dust extinction lanes.

Moving into the thermal IR reduces the total extinction,
mostly eliminates scattering by the dust and the UCHII regions become much
brighter. The interstellar dust grains are heated by a
combination of direct stellar radiation near the star and L$\alpha$
photons resonantly scattering in the ionized zone (e.g., {\em Natta and
Panagia}, 1976 and {\em Hoare et al.}, 1991). The latter process means
that the dust grains never drop below temperatures of a few hundred
Kelvin throughout the nebula and thus emit strongly in the mid-IR. As
such, the intrinsic mid-IR morphology is similar to that seen in
the radio, but again heavy extinction can intervene.

The advent of a new generation of mid-IR cameras on large telescopes
has yielded high quality mid-IR images of a number of UCHII regions.
In many cases the mid-IR and radio morphology are in good agreement.
Recent examples include W49A South ({\em Smith et al.}, 2000),
G29.96-0.02 ({\em De Buizer et al.}, 2002a), NGC6334F ({\em De Buizer
et al.}, 2002b), W3(OH) ({\em Stecklum et al.}, 2002), and K3-50A
({\em Okamoto et al.},  2003) and several sources observed by {\em
Kraemer et al.} (2003).  In the cometary G9.62+0.19B there are even
signs of a bright spot near the expected location of the exciting star
due to direct stellar heating ({\em De Buizer et al.}, 2003 and {\em
Linz et al.}, 2005).

There are other objects, however,  where a large extinction, even at 10
and 20~$\mu$m, greatly attenuates the mid-IR emission.  A
good example is G5.89-0.39, where the heavy extinction in the
near-IR obscures the southern half of the radio source in the N
and Q bands, consistent with the large column density measured in the
millimetre continuum ({\em Feldt et al.}, 1999). {\em De Buizer et
al.}  (2003, 2005) and {\em Linz et al.} (2005) show other instances
where the mid-IR morphology does not follow that of the radio
continuum. The most likely reason for this is extinction, which,
due to the silicate features, is still high in the ground-based mid-IR
windows. Indeed, the extinction at these wavelengths is only about
half of the value in the K-band ({\em Draine}, 2003).

Observations by the SPITZER satellite now offer a whole new IR
perspective on UCHII regions, in particular through the GLIMPSE survey
of a large fraction of the inner Galactic Plane ({\em Benjamin et
al.}, 2003).  The 4-8~$\mu$m range of the IRAC instrument has
extinction values lower than both the near- and ground-based mid-IR
windows, bottoming out around the 5.8~$\mu$m channel ({\em Indebetouw
et al.}, 2005). At these wavelengths one can expect to see some
emission from the hot grains in thermal equilibrium near the exciting
star. However, the IRAC filters at 3.6, 5.8 and 8.0~$\mu$m are
dominated by strong PAH features from the UCHII regions. The PAH
emission is strong in the photon-dominated regions (PDRs) that lie in
a thin shell of neutral gas just outside the ionization front. In
directions in which the H~II region is ionization bounded, the PAH
emission is expected to form a sheath around the nebula. If there are
directions where ionizing photons can escape, the PAHs will be
destroyed ({\em Girad et al.}, 1994).

\begin{figure*}
 \epsscale{2.0}
\plotone{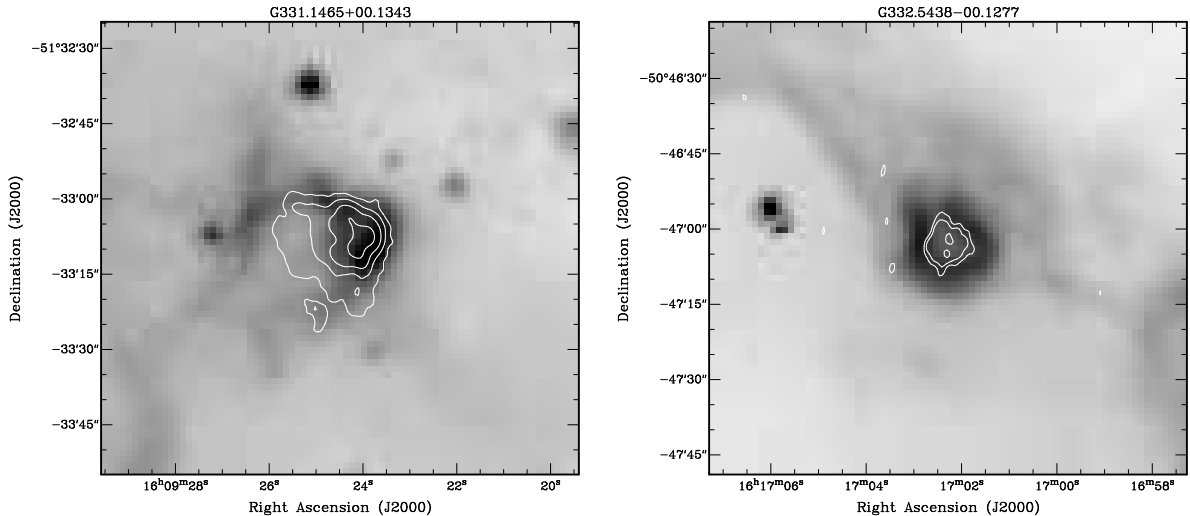}
\caption{\small Left: 8.0~$\mu$m GLIMPSE image of G331.1465+00.1343 
overlaid with contours of the 5 GHz radio continuum emission from ATCA 
with a noise level of 0.2~mJy per beam at a resolution of about 3~\arcsec\ 
The object has a clear cometary morphology seen in both
the radio and mid-IR image. 
Right: Same for G332.5438-00.1277. Note the 'horseshoe' shaped
IR emission that is open to the NNE similar to cometary objects and
not revealed by the barely resolved radio image.
For both sources there is a darker region ahead of the object in the 8.0~$\mu$m image
indicating extinction due to dense molecular cloud material
From {\em Busfield (2006)}, PhD thesis, University of Leeds.
}
\label{fig:cometfigs}
\end{figure*}

An example for a cometary UCHII region is shown
in Fig. \ref{fig:cometfigs} (left) where the 8~$\mu$m emission closely
follows the radio emission around the head and sides, but is absent in
the tail direction which is likely to be density bounded. Possibly unrelated
clumps and filaments can be seen in the vicinity, so caution must
be exercised when interpreting the SPITZER images; 
information from other wavelengths, especially the radio, is also needed.
Nevertheless, the uniformity and coverage of the GLIMPSE
dataset makes it very useful for studies of UCHII regions.
Its resolution and sensitivity can in principle yield new insights
into the morphologies of UCHII regions, notwithstanding the caveats
mentioned above.

{\em Hoare et al.}, (in prep.) are carrying out mid-IR morphological
classification utilising the GLIMPSE dataset.  They are using a sample
of massive young stars colour-selected from the lower resolution
mid-IR survey by the MSX satellite ({\em Lumsden et al.},  2002). These
have been followed-up with a variety of ground-based observations,
including high resolution radio continuum observations to identify the
UCHII regions and massive YSOs ({\em Hoare et al.}, 2004;
www.ast.leeds.ac.uk/RMS).  Preliminary findings indicate that the
proportion of cometary objects is higher than in previous radio
studies.  Fig. \ref{fig:cometfigs} (right) shows an example where
the radio image is barely resolved and yet the GLIMPSE 8~$\mu$m image
shows a 'horseshoe' shaped nebula wrapped around the radio source as
expected for PAHs in a PDR. However, the structure is open to the NNE
suggesting that the UCHII region is not ionization bounded in that
direction. Although it does not have the classic parabolic shape of a
cometary, it is suggestive of a champagne flow away from the dense
molecular cloud seen in extinction ahead of the object. Other data are
needed to confirm such an interpretation. A star is seen in the centre
of the nebula in the 3.6 and 4.5~$\mu$m GLIMPSE images and is likely to
be the exciting source. This demonstrates that the IRAC
wavelength range can address many aspects of UCHII regions.

%\bigskip
%\noindent
%\textbf{2.1.2 Extended Emission}
%\bigskip

{\em Kurtz et al.} (1999) found that many UCHII regions, when observed at lower
angular resolution, show extended, diffuse emission in addition to the
ultra-compact component.  The radio morphologies suggested a physical
connection between the UCHII region and the extended component for
a significant fraction of their sources.  {\em Kim and Koo} (2001), using a
different sample, made radio recombination line observations of UCHIIs
with extended emission.  The line velocities of the ultra-compact and
extended gas were nearly equal, supporting the idea of a physical
relationship between the two components.  {\em Ellingsen et al.} (2005)
searched for extended emission in eight southern UCHIIs with associated
methanol masers.  They found a lesser degree of extended emission than
reported for the random UCHII sample of {\em Kurtz et al.}  Methanol masers
may trace earlier stages of UCHII regions, hence the lesser degree of
extended emission in the {\em Ellingsen et al.} sample may reflect the relative
youth of those regions.

\begin{figure}[t]
\epsscale{1.0}
\plotone{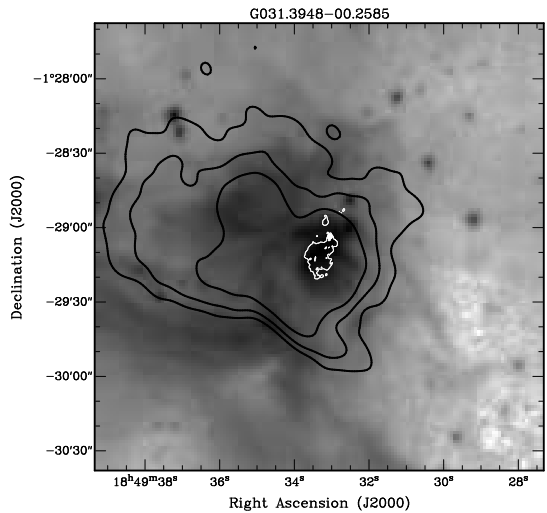}
\caption{\small GLIMPSE 8~$\mu$m image of G031.3948-00.2585 (IRAS
18469-0132).  Overlaid in thick black contours is the 3.6~cm data from VLA
D configuration observations and in thin white contours the VLA B
configuration data also at 3.6~cm. Radio data is from {\em Kurtz et
al.} (1999).  D array contours are at 1, 2 and 4~mJy per beam and then
switch to the B array contours at 0.5~mJy per beam up to the peak. The IR image
is more suggestive of a collection of separate UCHIIs and bright-rimmed clouds 
than a single source responsible for the compact and extended emission.}
\label{fig:G031.3948}
\end{figure}

As pointed out by
{\em Kurtz et al.} (1999) in many cases it is difficult to assess from
the radio data whether the extended emission is a coherent structure
excited by the same star(s). IR images can often clarify
the situation; the case of G031.3948-0.2585 (IRAS 18469-0132) is
shown in Fig. \ref{fig:G031.3948}.  The lower resolution radio map
of {\em Kurtz et al.} (1999) shows connected radio emission over two
arcminutes. In the IR image it looks more like a complex of
several different sources. The cometary, with a size of about
15\arcsec\ in the high resolution radio image, is seen along with IR
counterparts to the two radio point sources.  However, the diffuse
radio emission, seen extending to the east and north in the low resolution
radio map, appears to be due to a combination of another larger
cometary with its tail pointing north-east and an ionized
bright-rimmed cloud type structure running to the south and east of
the two cometary objects.  Spectrophotometric identification of the
individual exciting stars, together with high spatial and dynamic
radio continuum and line observations are needed to confirm the
picture in detail.

\bigskip

\noindent
\textbf{2.2 Molecular Environment}
\bigskip

In this section we examine the molecular environment within which
UCHII regions exist, and in particular, that of cometary
regions. Unfortunately, the typical spatial resolution of single-dish
millimetre line observations is usually insufficient to resolve UCHII
regions from associated star formation activity and the
molecular cores where they reside.  Higher resolution data on the
molecular environment can be obtained with
millimetre interferometry. However, most studies of the
environs of UCHII regions have naturally tended to concentrate on
those with associated hot cores. Although this benefits high mass
star formation studies in general, it is not so useful in the search
for a full understanding of UCHII region physics, a key ingredient of
which is the ambient density distribution.  The presence of a hot core
with its own associated infall and possibly outflow in close proximity
to the UCHII hampers a clear view of the molecular environment into
which the older UCHII region was born.

In single-dish studies {\em Hofner et al.} (2000) used optically thin
C$^{17}$O transitions to determine that the molecular clumps in which
UCHII regions reside are typically about 1 pc is size with densities
of 10$^{5}$~cm$^{-3}$ and temperatures of 25~K.  In their survey of
$^{13}$CO and CS transitions, {\em Kim and Koo} (2003)
found that the UCHII region is usually located right at the peak
of the molecular line emission. When they did resolve clear density
and velocity structure in the molecular gas, the pattern was consistent
with a champagne flow.

Somewhat higher resolution information on the density distribution is
available from sub-millimetre dust continuum emission maps ({\em
Mueller et al.}, 2002). These show very centrally condensed clouds
centred on the UCHII regions themselves, but if there is an offset of
the peak density then it is usually ahead of the cometary apex. Using
the sub-millimetre dust continuum will give a more centrally peaked
distribution than a pure column density tracer, as the warm dust in
and around the nebula will enhance the sub-millimetre emission.  {\em
Hatchell and van der Tak} (2003) examined the more extended
sub-millimetre continuum emission around UCHII regions and found that
the average radial density distribution derived was consistent with
the r$^{-1.5}$ expected for free-fall collapse.

About half of all UCHII regions are
associated with warm molecular gas as evidenced by highly excited
NH$_{3}$ ({\em Cesaroni et al.}, 1992), CS ({\em Olmi and Cesaroni}, 1999),
methanol masers ({\em Walsh et al.}, 1998) or other molecular tracers of the
early hot core phase ({\em Hatchell et al.}, 1998). Higher resolution data
showed that these hot cores were usually spatially offset from the
UCHII regions by a few arcseconds ({\em Cesaroni et al.}, 1994). Where these
are associated with cometary regions then the hot cores are always
located ahead of the apex.  Many cometary regions have
been found to show signs of earlier stages of massive star formation
such as maser sources or massive YSOs as well as hot cores in the
dense regions a few arcseconds ahead of the cometary region
(e.g., {\em Hofner et al.}, 1994; {\em Hofner and Churchwell}, 1996). 

G29.96-0.02 is again a good case study. Interferometric observations
of $^{13}$CO 1-0 and C$^{18}$O 1-0 by {\em Pratap et al.} (1999) and
{\em Olmi et al} (2003) reveal gas densities of
5$\times$10$^{5}$~cm$^{-3}$. The density peaks just in front of the
cometary H~II region, at the location of the hot
core. {\em Maxia et al.} (2001) found blue-shifted CO and HCO$^{+}$
1-0 emission in the tail region and HCO$^{+}$ 1-0 in absorption,
redshifted with respect to the main cloud at the head.  They
interpreted these motions as infall onto the hot core.

\begin{figure*}
 \epsscale{2.0}
\plottwo{hoare_fig3a.ps}{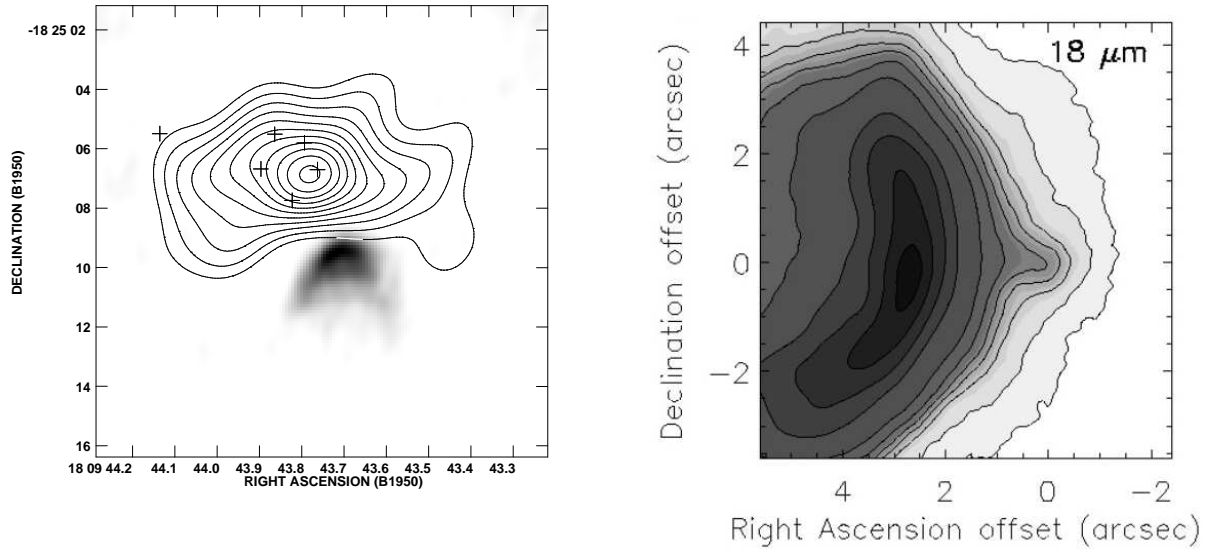}
  \caption{Left: High resolution map of the CS 2-1 emission (contours)
tracing the dense molecular gas ahead of the cometary UCHII region
G12.21-0.10 (radio continuum image in grey-scale). Crosses mark water masers. From {\em de la Fuente et al.}, in
preparation. Right: 18~$\mu$m image of G29.96-0.02 showing the detection of
the hot core located about 2 arcseconds in front of the cometary arc of the UCHII
region. From {\em De Buizer et al.} (2002a).}
\label{fig:G12.21}
\end{figure*}

Fig. \ref{fig:G12.21} shows a high resolution observation of the
dense molecular gas associated with the cometary region G12.21-0.10. 
The molecular gas lies just in front of the cometary ionisation front.
Observations of the NH$_{3}$ (2,2) line show that a hot core is
present, coincident with a collection of water masers showing
that star formation activity is occurring ({\em de la Fuente et
al.}, in prep.).  Recently these hot cores in the vicinity of UCHII
regions have been detected in the mid-IR via their warm dust emission
({\em De Buizer et al.}, 2002a, 2003 and {\em Linz et al.}, 2005).
Fig. \ref{fig:G12.21} shows the first mid-IR detection of such an
object ahead of the G29.96-0.02 cometary region.

In the few cases where a cometary object without obvious interference
from a nearby hot core or massive young stellar object has been
studied at high resolution, dense gas surrounding the head of the
cometary structure has been found.  Maps of W3(OH) in dense gas
tracers such C$^{18}$O and CH$_{3}$OH ({\em Wyrowski et
al.}, 1999) reveal an arc of emission wrapped around the western end
of the UCHII region, which has a champagne flow
to the east ({\em Keto et al.}, 1995).  Similarly, high resolution observations of G34.24 by
{\em Watt and Mundy} (1999) show C$^{18}$O 1-0 emission enveloping the head
of this cometary region, whilst the traditional hot core tracer,
methyl cyanide, peaks up right on the apex of the cometary
region. {\em Watt and Mundy} interpret this emission as arising from
dense molecular gas externally heated by the UCHII region rather than
internally heated by a young massive star. The interpretation of
molecular line emission close to the PDR around an UCHII region is
complicated by the many excitation and chemical effects that are
likely to be at work.

An alternative probe of dense molecular gas surrounding UCHII regions
is to use molecular lines in absorption against the strong continuum
background. This gives additional constraints due to the particular
geometry necessary to generate them. The H$_{2}$CO lines at 2 and 6 cm
have been observed at high resolution in absorption against W3(OH) and
W58 C1 ({\em Dickel and Goss}, 1987; {\em Dickel et al.}, 2001).  Both of
these objects have a cometary structure and the H$_{2}$CO line optical
depth maps of both objects show peaks consistent with the densest gas
being located at the head of the cometary region. Total gas densities of
6$\times 10^{4}$~cm$^{-3}$ were derived from the strengths of the
absorption lines. The authors interpret other details in the line
structure as arising from possible outflows from other sources in
these regions.

Very high spatial resolution absorption line studies have been
achieved with the NH$_{3}$ lines at 23 GHz. {\em Sollins et al.}
(2004) found evidence of spherical infall together with some rotation
in G10.6-0.4. A multi-transition study of G28.20-0.04 ({\em Sollins et
al.}, 2005) revealed complex motions interpreted as a combination of
toroidal infall and outflowing shell. These two H~II regions fall into
the hyper-compact category (see Section 3).

Lines that arise in the PDR can probe the conditions and
kinematics in the neutral material. Carbon radio
recombination lines are particularly useful as they can be
compared to the ionized gas traced by hydrogen and
helium recombination lines. {\em Garay et al.} (1998b) and {\em G\'{o}mez et
al.} (1998) observed these lines towards cometary objects and found
that the spatial distribution and velocity of the carbon lines are
consistent with an origin in a PDR around objects
undergoing a champagne flow.  The carbon lines are significantly
enhanced by stimulated emission and therefore arise
predominately from the near side in front of the continuum, and hence
give more specific insight into the velocity structure.  {\em Roshi et al.}
(2005a) used this fact to deduce that the PDR region in the cometary
object G35.20-1.74 is moving into the cloud at a few km~s$^{-1}$ by
comparison with the ambient molecular cloud velocity.  {\em Lebr\'{o}n
et al.} (2001) report a comprehensive study of a cometary region, 
G111.61+0.37, using a combination of atomic hydrogen 21 cm
emission and absorption components related to ionized and molecular
lines. They demonstrated that the atomic gas exhibits a champagne flow
like the ionized gas, but at significantly slower velocities, consistent
with entrainment. 

\bigskip

\noindent
\textbf{2.3 Stellar Winds}
\bigskip

If we are to understand the nature of UCHII regions we first need a
sound knowledge of the OB stars that excite them and their stellar
winds. Due to the heavy extinction, optical and UV
spectroscopic diagnostics are not available.  Even for the
objects that are visible in the near-IR when the extinction is low
enough it is usually difficult to see them because of the very strong
nebular continuum emission.  One of the few stars that is clearly
visible at 2~$\mu$m is the one that powers G29.96-0.02. {\em
Mart\'{i}n-Hern\'{a}ndez et al.} (2003) used the VLT to take an intermediate
resolution K-band spectrum of the star building on previous work by
{\em Watson et al.} (1997). By comparison with spectra of known spectral 
types from the field, they deduced that it was an O5-6V,
and argued that this was consistent with other estimates of
the spectral type. More importantly, the spectrum is consistent with
field stars which argues that the stellar wind should also have properties
similar to the well-studied radiation driven OB star winds. Hence, the
action of a strong wind moving at a few thousands of km~s$^{-1}$ has to
be an ingredient of any model of expanding UCHII regions.

The indirect information on the radiation output from the central
stars via the study of the excitation of the nebulae has been put on a
much firmer basis as a result of ISO spectroscopy of a large
sample. {\em Mart\'{i}n-Hern\'{a}ndez et al.} (2002) find a consistent
picture whereby the metallicity gradient in the Galaxy can explain the
excitation gradient due to the hardening of the stellar radiation
field. However, progress is still required on the stellar atmosphere
models to derive the exact stellar parameters.  The use of the near-IR
He I pure recombination lines can also help constrain the stellar
parameters ({\em Lumsden et al.}, 2003). It should now be possible to
combine the new spatially resolved mid-IR imaging with
multi-dimensional dusty photo-ionization modelling including PAHs to
determine the dust-to-gas ratio inside UCHII
regions. This is crucial to finding the fraction of Lyman continuum
photons absorbed by the dust which at present bedevils the use of
radio to IR luminosities in constraining the stellar effective
temperature.

There is an urgent need to find and conduct detailed direct studies of
the exciting stars from a range of UCHII regions. {\em Bik et al.}
(2005) have studied the spectra of many objects found in the vicinity
of UCHII regions. However these were either OB stars in a cluster
associated with, but not powering the UCHII region, or the UCHII
region itself where the spectrum is dominated by the nebular continuum
and not the star. To make progress, higher spatial resolution is
required to pick the star out of the nebular light. {\em Alvarez et
al.}  (2004) present a start down this road with near-IR adaptive
optics imaging of several UCHII regions. Higher Strehl ratios on
larger telescopes will be needed to carry out the high spatial and
spectral resolution observations to classify the exciting stars. The
speckle imaging of K3-50A by {\em Hoffmann et al.} (2004) shows the
promise of such techniques.

Another new line of evidence that can give insights into the stellar
winds in the centres of UCHII regions comes from X-ray
observations. Hard X-ray sources are now routinely found in high
mass star-forming regions. {\em Hofner et al.} (2002) made a
study of the W3 region using the {\em Chandra X-ray Observatory} and found
X-ray point sources coincident with many of the OB stars powering the
more evolved H~II regions, but also at the centres of some
ultra-compact regions. The point-like emission from the visible stars
in the evolved regions, which are not likely to be accreting, probably
arises in shocks within a radiatively driven stellar wind.
This could also be the case for the more compact
sources, but accretion shocks or wind-nebula interactions
are also possible. The diffuse X-ray emission arising from the very
hot post-shock region where the stellar wind interacts with the
surrounding UCHII region should fill the wind-blown cavity. 
Such diffuse emission is seen on much larger
scales where clusters of OB stars excite a large complex such
as M17 (see the chapter by {\em Feigelson et al.}). The indirect
effect of X-ray emission in UCHII regions is the generation of a
partially ionized layer just outside the nebula due to the
penetration of hard photons into the neutral material. {\em Garay et
al.} (1998a) observed narrow H I radio recombination line components,
which they ascribed to X-ray heating.

If the wind-nebula interaction can be detected in diffuse X-rays in
UCHII regions, it would in theory give an independent
measure of the strength of the wind. However, the modelling of this
interaction and its X-ray emission is far from straightforward.  The
evolution of wind driven  regions expanding into a high pressure
ambient medium was studied numerically by {\em Garc\'\i a-Segura and Franco}
(1996) who showed that the swept-up shells of ambient gas suffered a
strong thin-shell instability producing ``elephant-trunks'' and cometary globules
in the shocked stellar wind.  In these models, a hot
bubble of shocked stellar wind drives the expansion of a swept-up
shell where the ionization front is trapped.

{\em Comeron} (1997) made some initial calculations in the context of
stellar winds in a champagne flow scenario.  More recently,
{\em Gonz\'alez-Avil\'es et al.} (2005) studied the evolution of an H~II 
region driven by a strong stellar wind blowing inside a logatropic
cloud in gravitational collapse toward the central star.  They
included thermal conductivity which reduces the shocked stellar wind
cooling time by several orders of magnitude ({\em Shull}, 1980).  The
tangential magnetic fields will suppress the thermal conduction in the
direction perpendicular to the magnetic field direction. Nevertheless,
the morphology and strength of magnetic fields is unknown. Thus, the
magnitude of the thermal conductivity coefficient, $\kappa$, was used
as a free parameter.  They found that only models with the full
thermal value of $\kappa$ were compatible with the
hard X-ray luminosities observed in the W3 sources ($10^{-3}$~L$_{\sun}$)
by {\em Hofner et al.} (2002). For this value, thermal conductivity
cools the hot region of shocked stellar wind and the shells of
swept-up gas are driven by the momentum of the stellar wind. The hot
shocked wind is also in a very thin shell and would predict emission
on 0.1 pc scales that would easily be resolved by {\em Chandra}, which
was not the case for the W3 sources. Much deeper observations of this
and other massive star forming regions are being obtained and these
may reveal the diffuse X-ray emission that will fully unlock this
new insight into the effect of strong stellar winds in UCHII regions.

\bigskip

\noindent
\textbf{2.4 Dynamical Picture}
\bigskip

As reviewed by {\em Churchwell} (1999), many models have
been put forward to explain the morphologies and long lifetime
of UCHIIs.  {\em Wood and Churchwell} (1989b) were the first to argue
that the sheer numbers of objects with IR colours like UCHII regions
implies a lifetime of order 10$^{5}$~years rather than the
simple dynamical lifetime of 10$^{4}$~years. This has since been
criticised by several studies ({\em Codella et al.}, 1994; {\em Ramesh
and Sridharan}, 1997; {\em van der Walt}, 1997; {\em Bourke et al.},
2005), but nevertheless many models attempted to include ways of
lengthening the lifetime of the phase. These included
using thermal pressure to confine the expansion ({\em De Pree et al.},
1995), although this was claimed by {\em Xie et al.} (1996) to predict
too high an emission measure and they appealed to turbulent pressure
instead. Another way is to continuously introduce neutral
material to the ionized region. In mass-loading models by {\em Dyson
et al.} (1995), {\em Lizano et al. 1996} and {\em Redman et al.}
(1998) the ablation of neutral clumps sets up a recombination front
structure. These models predict mainly double-peaked line profiles
that are not observed. Magnetic fields may also impede the advance of
ionisation fronts ({\em Williams et al.}, 2000) and require sensitive
measurements of the field strength and geometry to test them.

For cometary objects, the bow shock model, whereby the stellar wind
from an OB star moving through a uniform medium sets up a parabolic
standing shock, also continuously feeds neutral material into the
nebula ({\em Mac Low et al.}, 1991). Several high resolution radio
recombination line studies of the dense gas at the head of cometary
regions, and G29.96-0.02 in particular, observed a velocity structure
that was claimed to support the bow shock model ({\em Wood and
Churchwell}, 1991 and {\em Afflerbach et al.}, 1994).  These pure bow
shock models required the star to be moving at velocities of about
20~km~s$^{-1}$ to impart sufficient velocity structure into the
gas. This is much higher than typical members of young OB star
clusters, which have velocity dispersions of only a few km~s$^{-1}$
(e.g., {\em Sicilia-Aguilar et al.}, 2005). Of course, there exists a
class of runaway OB stars that do have much higher proper motions,
probably resulting from dynamical ejections ({\em Hoogerwerf et al.},
2001 and see also {\em G\'{o}mez et al.}, 2005). However, these only
make up a small fraction of all OB stars ({\em Gies}, 1987; {\em de
Wit et al.}, 2005). The velocity dispersion of young clusters where
the potential is still dominated by gas will be higher. {\em Franco et
al.} (2006) have recently carried out numerical simulations of UCHII
regions in which the exciting star is moving with speeds of up to
13~km~s$^{-1}$ down a density gradient surrounding a hot core. They
identify these stellar speeds as being consistent with viral motion
due to the gravitational field of the hot core. Their motivation was
to explain the occurrence of both compact and extended emission if due
to a single ionising source (see Section 2.1.2).

Lower spatial resolution radio recombination line studies (e.g., {\em
Garay et al.}, 1994; {\em Keto et al.}, 1995) that are sensitive to
more extended emission in the tail of cometary objects, indicate
flowing motions consistent with the blister and champagne flow
models ({\em Israel}, 1978; {\em Yorke et al.}, 1983).  Near-IR
long-slit spectroscopy also showed that the velocity structure in the
Br$\gamma$ line in G29.96-0.02 was not consistent with a pure
bow shock picture ({\em Lumsden and Hoare}, 1996, 1999). The same
velocity structure was recently recovered by {\em Zhu et al.} (2005)
using long-slit echelle observations of the 12.8~$\mu$m [Ne II] fine
structure line.  A high quality near-IR spectrum was presented by {\em
Mart\'{i}n-Hern\'{a}ndez et al.}  (2003) which allowed a more accurate
comparison with the surrounding molecular gas through the presence of
the H$_{2}$ S(1) line. Its velocity agreed well with millimetre
molecular line measurements even though it is likely to arise in a
thin layer just outside the ionized zone.

Comparison with the molecular cloud velocity is a vital part of
testing dynamical models since it sets the reference frame. For
instance, in G29.96-0.02, where the tail
is clearly pointing towards us, the bow shock model would predict that
most of the gas is redshifted as the stellar wind shock tunnels into
the molecular cloud. At the other extreme, a pure champagne flow
scenario would involve most of the gas being blueshifted as it
photo-evaporates off the molecular cloud face and flows back down the
density gradient towards the observer. In G29.96-0.02, and other
simple cometary objects where sufficiently good spatially resolved
velocity data for the ionized and molecular gas exists (e.g., {\em Keto
et al.}, 1995; {\em Garay et al.}, 1994; {\em Hoare et al.}, 2003;
{\em Cyganowski et al.}, 2003), the range of velocity structure
usually straddles the molecular cloud velocity.

As discussed in Section 2.3, stellar winds must be 
included in any realistic model of expanding H~II regions. The limb-brightened
morphology in the ionized gas of the shell and cometary regions gives
the appearance that the central regions are cleared out by a strong
stellar wind. Pure champagne flow models ({\em Yorke et al.}, 1983)
have a centrally peaked radio structure. Numerical modelling by {\em
Comeron} (1997) first indicated that winds can produce limb-brightened
cometary structures.  The action of the wind also helps to slow the
expansion, as the ionization front is trapped in the dense swept up
shell, as first pointed out by {\em Turner and Matthews} (1984).  The
spherically symmetric wind-driven H~II region models by {\em
Gonz\'alez-Aviles et al.}  (2005) produced very thin shell nebular
regions. They found that when the H~II region is expanding into a
logatropic density distribution $\rho \propto r^{-1}$, driven by
stellar winds with mass-loss rates $\dot M_w \sim 10^{-5}$~M$_{\sun}
yr^{-1}$ and wind speeds, $v_w \sim 1000$~km~s$^{-1}$, the H~II shells
evolve from HCHII to UCHII sizes in timescales $ \Delta t \sim 3 - 7
\times 10^4 \,$~yr.

\begin{figure*}
 \epsscale{2.0}
\plotone{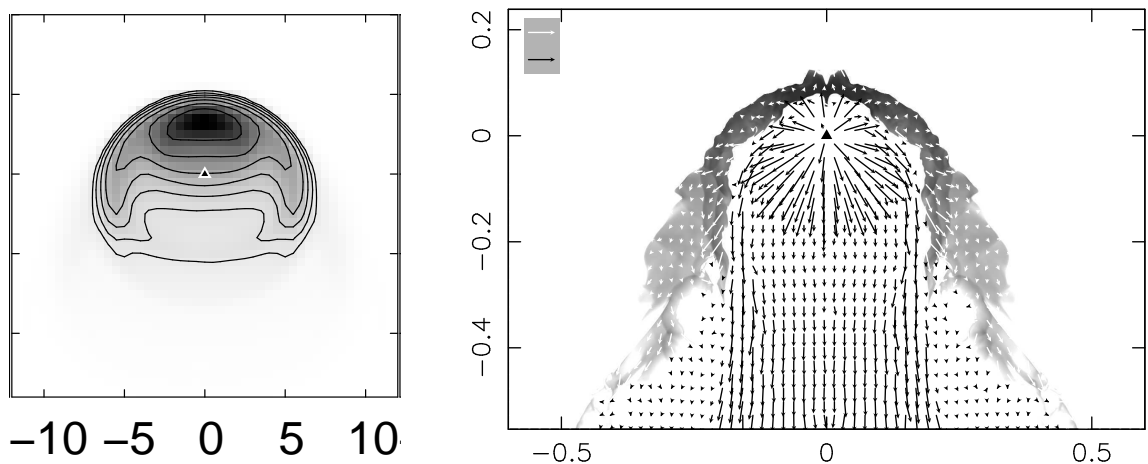}
\caption{\small Results from hydrodynamic modelling of a cometary
UCHII region expanding into an exponential density gradient of scale
height 0.2 pc excited by a star with stellar wind typical for an early
type star and moving up the density gradient by 10~km~s$^{-1}$. Left:
Predicted emission measure map at age 20 000 years with size scale in
arcseconds for an adopted distance of 7 kpc. Right: Corresponding
velocity structure of the model with white arrows showing velocities
up to 30~km~s$^{-1}$ and black arrows up to 2000~km~s$^{-1}$. The
ionisation front at the head moves up the density gradient at about
3~km~s$^{-1}$. From {\em Arthur and Hoare} (2006).}
\label{fig:hydro}
\end{figure*}

Recently, new hydrodynamical models of cometary H~II regions by {\em
Arthur and Hoare} (2006) considered the action of fast
($\sim$2000~km~s$^{-1}$), strong (\mdot $\sim$ 10$^{-6}$ M$_{\sun}$)
winds, typical for early type stars, within steep power-law and
exponential density gradients.  Fig. \ref{fig:hydro} (left) shows the
emission measure map from one such model that has a limb-brightened
structure very much like those seen in cometary objects. If
the mass-loss rate is dropped by an order of magnitude then the wind
can no longer open up a significant cavity.

\begin{figure}[t]
 \epsscale{1.0} \plotone{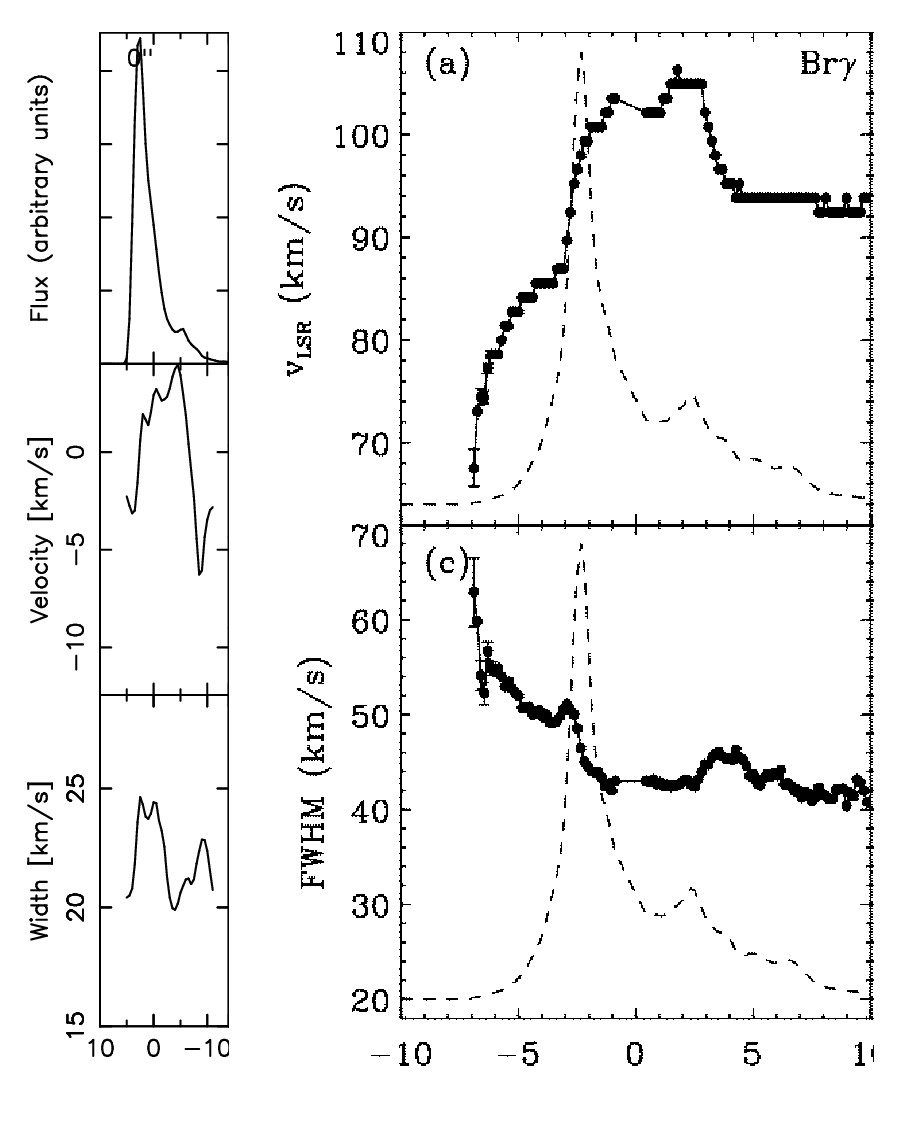}
\caption{\small Left: Simulated long-slit spectral data along the axis of
the model cometary H~II region shown in Fig. \ref{fig:hydro}. The
three panels show the flux, velocity centroid and line width respectively.
Velocities are with respect to
the molecular cloud which is at rest in the simulation.
Horizontal axis is offset from the position of the exciting star in arcseconds.
Right: Long-slit spectral data
for a position approximately along the axis of G29.96-0.02 from {\em Mart\'{i}n-Hern\'{a}ndez
et al.} (2003). Top panel shows the velocity centroid measured for the
Br$\gamma$ line whilst the bottom shows the observed line width with the flux profile in the dotted line.
The best estimate of the ambient molecular cloud velocity is $v_{LSR}$=98~km~s$^{-1}$.
Horizontal axis is offset from the exciting star in arcseconds.}
\label{fig:janefig2}
\end{figure}

In these champagne models with a stellar wind, the ionized gas is
forced to flow in a parabolic shell around the bubble blown by the
stellar wind and then on down the density gradient.  To achieve the
right velocity structure relative to the molecular cloud a bow shock
element is also required, whereby the star is moving up the density
gradient. A stellar speed in the range 5-10~km~s$^{-1}$ is sufficient
to provide enough forward motion of the gas in the head to shift those
velocities relative to the molecular cloud by the typical amount
observed. Fig. \ref{fig:hydro} (right) shows the velocity structure
for such a combined champagne and bow shock model. This model produces
a velocity structure similar to that seen in G29.96-0.02 (Fig.
\ref{fig:janefig2}). Such a hybrid between pure champagne and pure bow
shock models was also developed by {\em Cyganowski et al.} (2003) to
explain the twin cometary regions in the same molecular cloud in
DR21. The lifetimes predicted by such models are determined by how fast the
ionisation front or the star reaches the local density maximum. For
G29.96-0.02 and the DR21 regions the density maximum is of order 0.1
pc ahead of the ionisation front. For ionisation fronts advancing at
about 3~km~s$^{-1}$ as in the {\em Arthur and Hoare} (2006) models this
gives a lifetime of a few times 10$^{4}$~years, similar to the
timescale for the star to reach the same point. {\em Kawamura and
Masson} (1998) measured the expansion speed of the cometary
W3(OH) region to be 3-5~km~s$^{-1}$.

{\em Zhu et al.} (2005) developed a model of flow around a parabolic
shell for the G29.96-0.02 and Mon R2 cometary regions including
pressure gradient forces resulting from the sweeping up of material by
the bow shock, but still within a constant ambient density.  As with
the original bow shock models for G29.96-0.02, this required a stellar
velocity of 20~km~s$^{-1}$ resulting in
a mis-match with the molecular cloud reference frame of
10~km~s$^{-1}$.  {\em Lumsden and Hoare} (1996, 1999) also needed to
invoke a 10~km~s$^{-1}$ shift in the opposite direction to get their
semi-empirical champagne flow model to agree with the observations,
which they ascribed to expansion of the ionization front.  Overall, the
picture that is emerging for the few well-studied cometary regions is
that the morphology, velocity structure and lifetime of cometary UCHII
regions can be explained by a combination of champagne flow down a
density gradient and bow shock motion up the density gradient. This
still remains to be tested on a wider range of objects. Possible
explanations of the direction of motion will be discussed by {\em
Hoare et al.}, in preparation.

\bigskip
\centerline{\textbf{ 3. HYPER-COMPACT H~II REGIONS}}
\bigskip

\noindent
\textbf{3.1 Observed Properties}
\bigskip

Within the large number of UCHII regions that have been discovered, a
number of objects stand out as being exceptionally small and
dense. Sizes for this group are $\leq 0.05$~pc (10,000 AU) while
densities are $\geq 10^6$~cm$^{-3}$ and emission measures are $\geq
10^{10}$~pc~cm$^{-6}$.  In the past few years, these regions have come
to be considered as a separate class, referred to as {\em
hyper-compact} H~II regions (HCHII).  These regions are considered a
distinct class from UCHII regions --- rather than merely representing
the most extreme sizes and densities --- primarily because they have
extremely broad radio recombination line profiles, with $\Delta$V
typically 40 -- 50~km~s$^{-1}$, and some greater than 100~km~s$^{-1}$
({\em Gaume et al.}, 1995; {\em Johnson et al.}, 1998; {\em Sewilo et
al.}, 2004).  By comparison, UCHII regions typically have
recombination line widths of 30 -- 40~km~s$^{-1}$ ({\em Keto et al.},
1995; {\em Afflerbach et al.}, 1996).  Although the precise
relationship between HCHII regions and Broad Recombination Line
Objects (BRLO) is not yet clear, at present the two classes have so
many objects in common that we take them to be one and the same.

There are also similarities between the HCHII/BRLOs
and the massive YSOs that have strong ionized stellar winds, including
BN, W33A, Cep A2, GGD27 and S140 IRS 1 ({\em Rodr\'{i}guez}, 1999;
{\em Hoare}, 2002). These luminous ($>10^{4}$~L$_{\sun}$), embedded
sources are often found without hot molecular core signatures, but are usually
still driving outflows and probably accreting. We do not use the
term ``protostar'', since it is likely that the luminosity of
these objects is dominated by hydrogen burning and not accretion,
which cannot be observationally distinguished at present.

The massive YSOs are weaker radio sources and have
therefore not been the subject of radio recombination line
studies. However, their IR recombination line profiles have been
studied and are very broad, usually in excess of 100~km~s$^{-1}$
(e.g., {\em Bunn et al.},  1995). High resolution radio mapping has
yielded both jet and equatorial wind morphologies for the nearest
sources ({\em Hoare}, 2002; {\em Patel et al.}, 2005). Proper motion
studies of jet sources reveal velocities of 500~km~s$^{-1}$ ({\em
Mart\'{i} et al.}, 1998) and the high collimation suggests a
magneto-hydrodynamic mechanism as in low-mass YSOs.  (Note
the ionized X-wind or 'outflow-confined' model put forward for HCHII
regions by {\em Tan and McKee} (2003) is more appropriate for these
jet sources than the slower HCHII regions.)  The equatorial wind sources can
result from the pressure of stellar radiation acting on the gas on the
surface of the accretion disk ({\em Sim et al.}, 2005). 

A key difference between wind sources and UCHII regions is that in
wind sources the ionized material originates from the star-disk system
itself. In contrast, for expanding UCHII regions it is the surrounding
molecular cloud material that is being ionized.

\begin{figure*}[t]
 \epsscale{2.0}
\plottwo{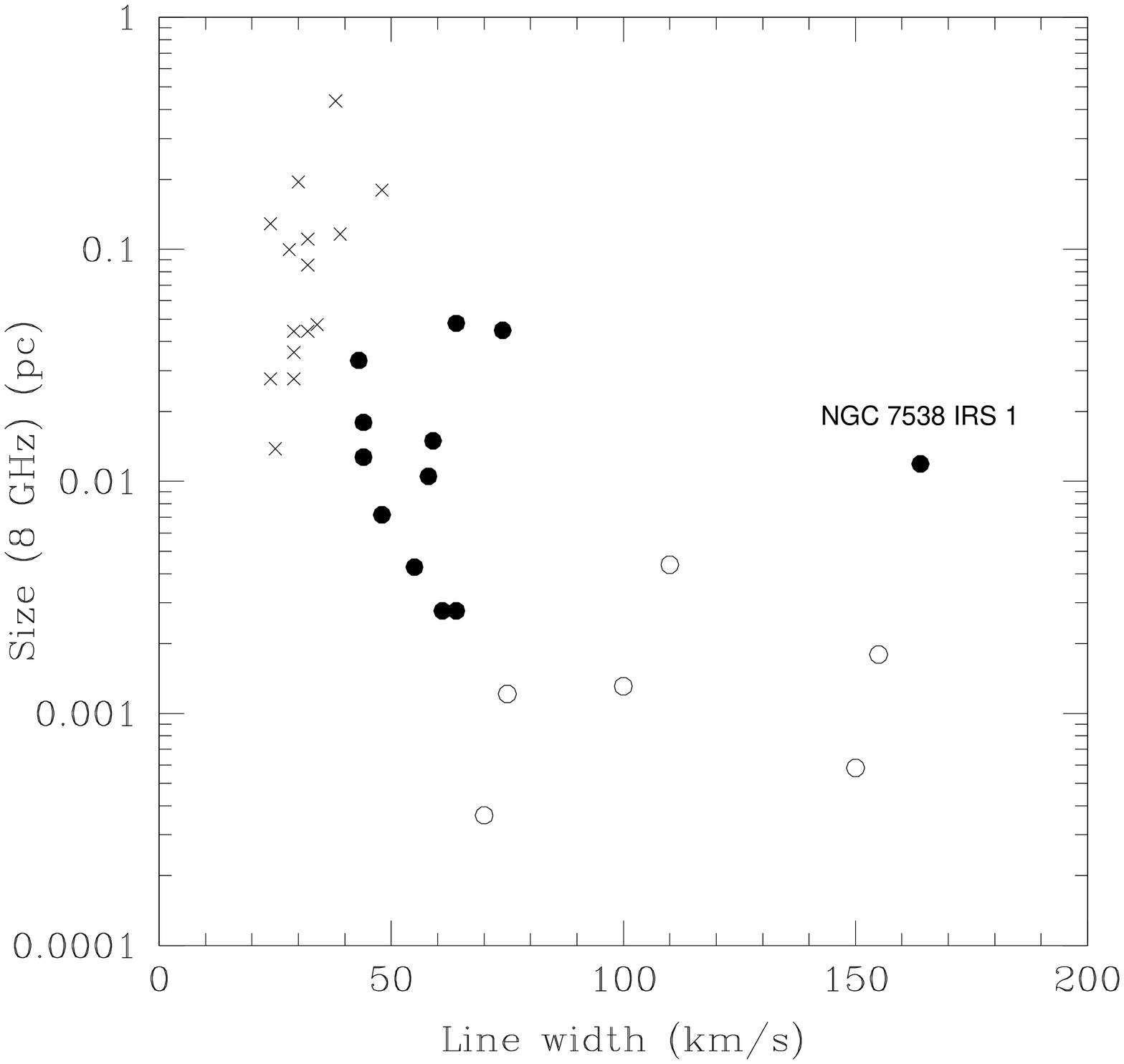}{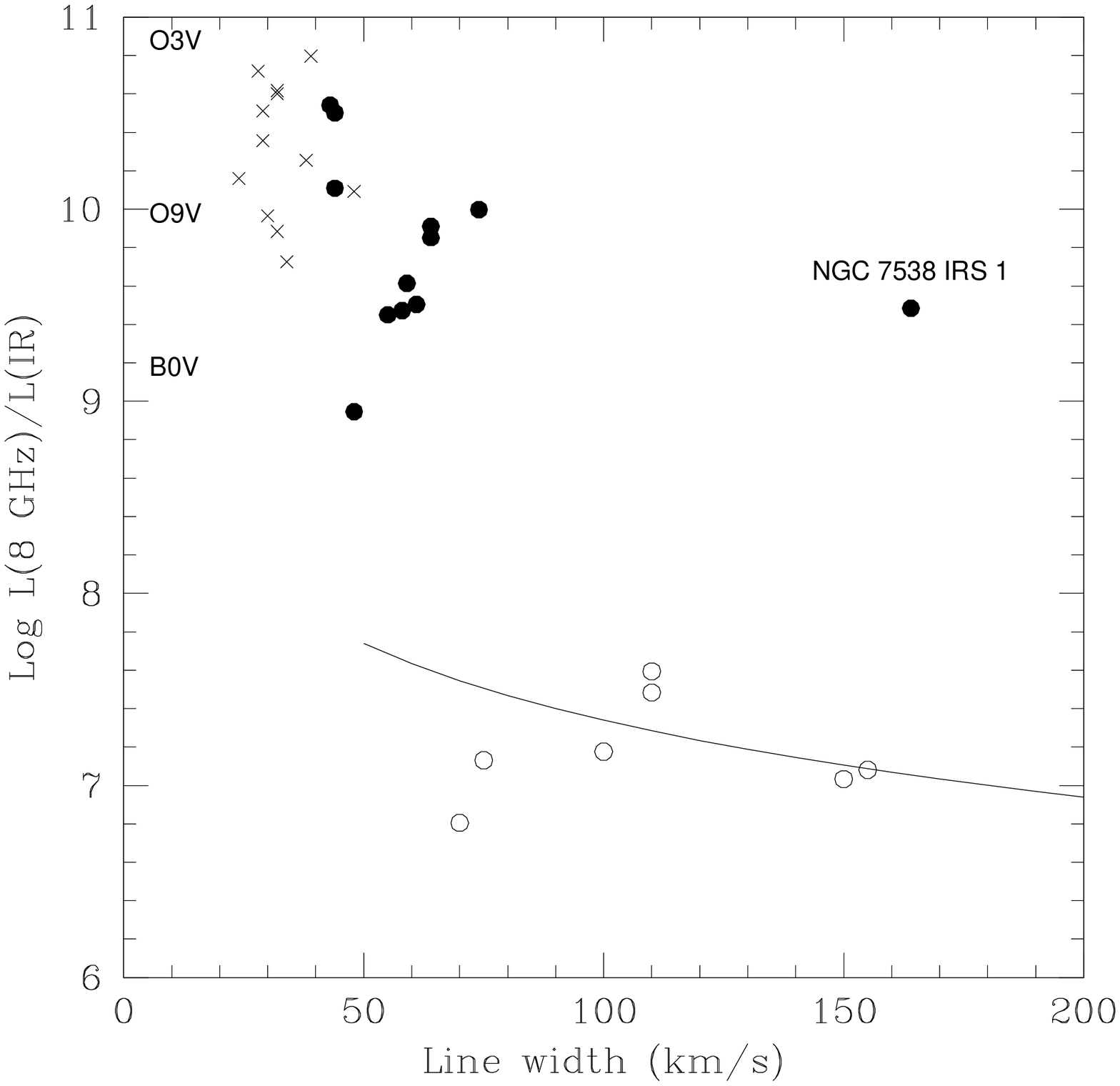}
  \caption{Left: Size versus line width for UCHII regions (crosses), HCHII
regions (solid circles) and massive young stellar object wind sources
(open circles).  Line widths are FWHM. Sizes are measured or equivalent at 8~GHz and are geometric
mean FWHM. Note how the HCHII regions lie between
UCHII regions and the MYSOs wind sources. 
Right: As left, but for the ratio of the radio luminosity at 
8~GHz (W~Hz$^{-1}$) to the bolometric luminosity
from the IR (L$_{\sun}$) (except W49A regions where radio spectral type has been used). 
The flux at 8 GHz has been
extrapolated from higher frequencies if not measured directly, using
observed or typical spectral indices. Data for
UCHIIs are from {\em Wood and Churchwell} (1989a) and {\em De Pree et al.} (2004); for the HCHII
regions from {\em Sewilo et al.} (2004), {\em Jaffe and Martin-Pintado} (1999), {\em Johnson et
al.} (1998), and {\em De Pree et al.} (2004); and for the MYSO wind sources from
{\em Bunn et al.} (1995) and {\em Nisini et al.} (1994). Sizes of MYSOs come from {\em Hoare} (2002)
and references therein; {\em Rengarajan and Ho} (1996); {\em Snell and Bally} (1986); {\em Campbell et al.} (1986).
The expected ratio for optically thin H~II regions for given spectral type exciting stars is indicated
at top left using stellar parameters from {\em Smith et al.} (2002). The solid line shows
the level expected for a stellar wind assuming $\mdot$=10$^{-6}$~M$_{\sun}$~yr$^{-1}$, $v_{\inf}=2v_{FWHM}$
and L(bol)=10$^{4}$~L$_{\sun}$ ({\em Wright and Barlow}, 1975).}
\label{fig:types}
\end{figure*}

The observed properties of HCHII regions would appear to be
intermediate between these two extremes, as illustrated in
Fig. \ref{fig:types}.  The size versus line width plot is an extension
of previous plots, e.g., {\em Garay and Lizano} (1999).  The line
widths for the MYSO wind sources are from the near-IR Br$\gamma$ line,
whilst those for the UCHII and HCHII regions are from radio
recombination lines. The highest frequency data available was used to
minimize the effects of pressure broadening. For winds, the size
varies with frequency and the measured or extrapolated size at 8 GHz
is plotted.  The UCHII region size does not vary with frequency and we
assume the same for HCHII regions, although they may vary somewhat.

For the right-hand plot of the radio brightness normalised to the
total luminosity, the HCHII and UCHII regions are clearly much more
radio-loud than the MYSO winds.  The 8~GHz radio luminosity is not a
straightforward property to use, because it under-estimates the number
of ionizing photons produced by the exciting source when the source is
optically thick. At 8~GHz, HCHII regions in particular, are well below
their turnover frequency, and present correspondingly lower flux
densities.  There is also a strong dependence on the effective
temperature of the ionizing source. It should be noted that the MYSO
wind sources plotted (BN, W33A, S140 IRS 1, NGC 2024 IRS 2, GL 989 and
GL 490) are somewhat lower luminosity than most of the H II regions
plotted. However, the combination of these observable parameters
provides a reasonable indication as to what type of object one is
dealing with.  It is interesting to note that NGC 7538 IRS 1 stands
out in these plots as an exceptional source due to its broad lines and
strong emission.

Two different physical mechanisms may contribute to the broad
line-widths of HCHII regions.  Pressure broadening can be significant
in such high density regions, particularly for the high principle
quantum number centimeter wave transitions observed with the VLA,
owing to the $n^7$ dependence of the broadening (e.g., {\em Brocklehurst and
Seaton}, 1972; {\em Griem}, 1974; see also {\em Keto et al.}, 1995 for a discussion of
the combination of dynamics and line broadening as applied to H~II
regions).  Bulk motion of the gas, either via accretion (e.g., {\em Keto},
2002b, {\em Keto and Wood}, 2006), outflow (e.g., {\em Lugo et al.}, 2004), or
rotation, could be occurring.  To distinguish the relative
contributions from pressure broadening and bulk motion of the gas will
require high spatial resolution ($\leq 1''$) observations over a wide
range of frequencies, including low principal quantum number
transitions in the millimeter, sub-millimeter and IR regions.

Apart from the common occurrence of broad recombination lines, 
other differences exist between the hyper- and ultra-compact classes.
UCHII regions typically have turnover frequencies (between the
optically thick and thin regimes) of 10 -- 15~GHz, while
HCHII regions become optically thin at frequencies above 30~GHz,
owing to their higher densities.  Nevertheless, no HCHII
regions have been found with an optically thick spectral index of +2,
indicative of uniform density gas.  Rather, HCHII regions show an
intermediate spectral index of $\sim +1$ which suggests non-uniform
gas density. Some UCHII regions also show $\alpha \sim +1$ at
centimeter wavelengths. Density gradients have been invoked to explain
these intermediate spectral indices (e.g., {\em Franco et al.}, 2000;
{\em Avalos et al.}, 2005).  Although a density gradient and small
size are suggestive of a stellar wind, the measured flux densities and
the inferred electron densities are much higher than values
encountered in stellar winds.  A disk wind, however, may be capable of
producing a density gradient, and at the high electron
densities implied by the radio flux densities ({\em Lugo et al.},
2004; see \S 3.2).  The effect of champagne flows or influence of
stellar gravity (Keto 2003) lead to steep density gradients.  {\em
Ignace and Churchwell} (2004) have shown that the intermediate
spectral indices can also be produced by unresolved clumps with a
power-law distribution of optical depths.

Morphologies of HCHII regions are not well-known; to date, only a few
have been resolved by radio continuum observations.  The archetypal
BRLO NGC 7538 IRS 1 is bipolar, but other morphologies are seen as
well.  Fig. \ref{fig:HCHII} shows radio continuum images of NGC 7538 IRS 1,
M17-UC1, and G28.20$-$0.04 which have, respectively, bipolar,
cometary and shell-like morphologies. The high spatial resolution
observations needed to identify the nature of the broad recombination
line emission will also provide much-needed morphological information.

Although classifications based on observed parameters can be useful, a
more fundamental definition relating to the nature of the object is
preferable.  One possibility is to define HCHIIs as those regions with
radii less than the transonic point where the escape velocity from the
star(s) within the H~II region equals the ionized gas sound speed (see
\S 3.2).  Alternatively, one might consider the place of HCHIIs in the
general massive star formation process.  In this context, the crucial
element is the recognition that HCHII sizes are 10,000 AU {\em and
smaller}.  Regardless of the model of massive star formation that is
adopted, it is clear that HCHII sizes are approaching a scale
appropriate for individual high mass star formation.  

\bigskip

\noindent
\textbf{3.2 Theoretical Models}
\bigskip

\noindent
{\it 3.2.1 Context - Hot Molecular Cores} The observed properties of
the HCHII regions suggest that they are very young and it is natural
to associate them with the turn-on of the Lyman continuum radiation.
H~II regions around stars later than early-B type are not detectable
by present radio telescopes.  Thus, if a star is growing by accretion
through the spectral type sequence from B to O (see \S3.2.2), then in
the earlier stages the star and accretion flow would appear as a hot
molecular core (HMC) without radio continuum emission.  {\em Osorio et
al.}  (1999) modelled the spectral energy distribution of the thermal
emission from dust in such HMCs accreting toward a central star.  They
modelled the HMCs as logatropic cores with pressure $P = P_0
\log(\rho/\rho_{{\rm ref}})$, where $P_0 \sim 10^{-7}-
10^{-6}$~dynes~cm$^{-2}$ is the pressure constant and $\rho_{{\rm
ref}}$ is a reference density ({\em Lizano and Shu}, 1989; {\em Myers
and Fuller}, 1992).  Logatropic cores have an equilibrium density
profile $\rho \propto r^{-1}$, and a large velocity dispersion
$\sigma^2 = dP/d\rho = P_0/ \rho,$ that increases with decreasing
density, as observed in molecular clouds (e.g., {\em Fuller and
Myers}, 1992) supporting massive envelopes of several hundred
M$_\sun$.  The gravitational collapse of logatropic spheres has the
property that the mass accretion rate increases with time as $\dot M
\propto t^3$ ({\em McLaughlin and Pudritz} 1997).  In a comparison of
their model with observations of several hot molecular cores without
detectable radio continuum {\em Osorio et al.} found that the observed spectral
energy distributions (SEDs) were consistent with central B stars with
ages $\tau_{\rm age} \ltappeq 6 \times 10^{4}$ yr, accreting at rates
of $\dot M_{\rm acc} > 10^{-4} \,M_\odot \, {\rm yr}^{-1}$.  At these
accretion rates, the main heating agent is the luminosity arising from
the deceleration of the accretion flow, $L_{\rm acc} = G M_\ast \dot
M_{\rm acc} / R_\ast$ (assuming purely spherical accretion).
Comparison of observed and modeled SEDs suggests that some HMCs are
consistent with the earlier stages of the formation of massive stars
by accretion.

The logatropic core model is also attractive because it has
properties that are required if massive stars are to form despite the
constraints imposed by the outward force of radiation pressure and
within the short time scale allowed by the main sequence lifetime.
{\em Osorio et al.} showed that their inferred accretion rates
were large enough that the momentum of the accretion flow could
overcome the outward force of radiation pressure on dust grains.  At
the risk of over-simplicity, this argument may be summarised as
follows:  The most naive estimate of the outward force deriving from
the luminosity, assuming spherical geometry and the total absorption of
the luminosity by the dust in the flow, would be $L/c$.
%\sim 1\times 10^{29} (L / 7.5 \times 10^5 L_{\sun})$ dynes.  
A similarly simple
estimate of the force deriving from the momentum of the accretion flow,
again assuming spherical accretion, is $\dot {\rm M} v$. If the
momentum of the accretion flow is high enough, the flow will push the
dust grains inward until they are sublimated by the higher temperatures
in the center of the flow. At this point  the flow is rendered
essentially transparent to the stellar radiation ({\em Kahn}, 1974; {\em Wolfire and
Cassinelli}, 1987).

\bigskip
\noindent
{\it 3.2.2 Ionized Accretion Flow or Gravitationally Trapped H~II regions} 
Once the star, possibly one of the several stars forming together, has
gained sufficient mass and therefore temperature, the number of emitted
Lyman continuum photons will be enough to ionize an H~II region
within the continuing accretion flow. The radius of ionization
equilibrium, $r_i$, is set by the balance between ionization and
recombination within the H~II region.  If $r_i$ is less than the 
gravitational radius in the accretion flow, $r_g \equiv G M_\star / a^2$,
where the inward velocity equals the sound speed
of the ionized gas, $a$ (within a factor of 2), then the H~II region can be
trapped within the accretion flow with its boundary as a stationary
R-type ionization front.  This model explains why the development of an
HII region does not immediately end the accretion as would be expected
in the classic textbook model for the evolution of H~II regions by
pressure driven expansion ({\em Shu}, 1992; {\em Spitzer}, 1978; {\em Dyson and Williams}
1980).  In pressure driven expansion, the flow of the ionized gas is
entirely outward, precluding accretion.  The model for H~II regions
trapped within an accretion flow was developed to explain the
observations of the inward flow of ionized gas toward the massive stars
forming in the H~II region G10.6--0.4 ({\em Keto}, 2002a).

The accretion model assumed in {\em Keto} (2002b) is a spherical steady-state
flow (Bondi accretion) which has a density gradient everywhere less
steep than $n\sim r^{-3/2}$. This is significant in that the spherical
solution for ionization equilibrium does not allow for solutions in
steeper density gradients ({\em Franco, Tenorio-Tagle and Bodenheimer}, 1990).
Similar to accretion in a logatropic sphere, 
Bondi accretion has the property that the accretion rate depends on
the stellar mass (as M$^2$ in Bondi accretion). As in the model of
{\em Osorio et al.} (1999), this model relies on the momentum
of a massive accretion flow to overcome the radiation pressure of the
stars. Observations of G10.6--0.4 indicate that the outward force from
the observed luminosity of $1.2\times 10^6$~L$_\odot$ ({\em Fazio  et al.},
1978) is approximately equal to the inward force deriving from the
momentum of the flow, $\dot {\rm M} v \sim 3\times 10^{28}$~dynes, for
an estimated accretion rate of $10^{-3}$~M$_\odot$ yr$^{-1}$ and
velocity of 4.5~km~s$^{-1}$ at 5000~AU ({\em Keto}, 2002a).  In this model,
the accretion flow, if not reversed by radiation pressure, will end
when the ionization rate, which increases with the increasing mass
of the star(s), is high enough that the radius of ionization
equilibrium increases beyond the gravitational radius.

While the central star is accreting mass at rates $\dot M \sim
10^{-3}$ ~M$_\odot {\rm yr}^{-1}$, the ram pressure of the accreting
material will be larger than the ram pressure of the stellar wind,
$\rho_{\rm acc} v^2_{\rm acc} >> \rho_{\rm w} v^2_{\rm w}$, and will
prevent the stellar wind from blowing out.  If the radiation pressure
on dust grains reverses the accretion flow, or the HCHII region
expands because its radius becomes larger than $r_g$, then a stellar
wind can blow out (see also {\em Gonz\'alez-Avil\'es et al.}, 2005).

Recently, {\em Keto and Wood} (2006) extended the model of {\em Keto}
(2002b) to consider accretion flows with angular momentum as described
by {\em Ulrich} (1976) and {\em Terebey et al.} (1984). In this model
of accretion, the gas spirals in to the star on ballistic trajectories
conserving angular momentum. An accretion disk develops at a radius
$r_D$, where the centrifugal force balances the gravitational force,
$\Gamma^2/r_D^3 = GM/r^2_D$, roughly where the infall velocity equals
the rotational velocity. Here $\Gamma$ is the initial specific angular
momentum. Since $r_D \sim GM/v_{orbital}^2$, if $v_{orbital} < c$
at the gravitational radius, $r_g$, then the radius of disk formation
will be within the maximum radius, $r_g$, of a trapped H~II
region. Thus the radius of disk formation, depending on the flux of
ionizing photons, could be within the ionized portion of the flow,
meaning that the accretion disk would form out of infalling
ionized gas. Note that the central parts of the centrifugal disk
should be neutral because the gas recombines at the high densities
expected in these disks.  With a different choice of parameters,
specifically, if the ionizing photon flux or the angular momentum were
greater than in the previous case, then the disk might form in the
molecular portion of the flow.  In this case, the accretion flow would
have the same structure as that assumed in the model of
photo-evaporating disks of {\em Hollenbach et al.} (1994) and {\em
Johnstone et al.} (1998).

{\em Keto and Wood} (2006) proposed the following evolutionary hypothesis for
the development of an H~II region within an accretion flow with a disk
that parallels the development of an H~II region within a spherical
accretion flow.  In the initial stage when there is no H~II region,
necessarily $r_D > r_{i}$ and the flow is described by the
massive molecular accretion disk. In the second stage, a trapped
HCHII region will develop in the center of the disk.
Because the gas in the disk is denser than the gas elsewhere around
the star, the molecular accretion disk will not necessarily be fully
ionized, but because $r_{i} < r_g$, the ionized surface of the disk
will not be expanding off the disk. Rather there will be a limited
region, contained within the H~II region, with an ionized accretion
flow onto the disk.  However, depending on the initial angular
momentum in the flow, this region may be very small with respect to
the extent of the molecular disk. Outside this region, there will be a
molecular accretion flow onto the disk.  The third stage of evolution
is defined by the condition $r_{ionized} > r_G$. In the non-spherical
case, because the gas density around the star is a function of angle
off the disk, the ionization radius, $r_{i}$ will also be a
function of angle. If the disk is sharply defined as in the
{\em Ulrich} (1976) and {\em Terebey et al.} (1984) models, then in the
third stage the H~II region will expand around the disk. In this third
stage, because $r_{i} > r_g$, the surface of the disk 
(except for a small region in the center) will photo-evaporate
with an outward flow of ionized gas off the disk, as described in the
models for photo-evaporating disks.

\bigskip
\noindent
{\it 3.2.3 Photoevaporating Disks}
{\em Hollenbach et al.} (1994) and {\em Johnstone et al.} (1998)
proposed that unresolved UCHII (now termed HCHII regions)
arise from photo-evaporating molecular disks.  The photo-evaporation
of circumstellar molecular disks around massive stars occurs as the
disk surface is ionized by the stellar Lyman continuum photons.  {\em
Hollenbach et al.} proposed that within the gravitational
radius, $r_g $, the heated gas is confined in the gravitational
potential well of the star.  For $r > r_g$ the ionized gas can escape
and an isothermal evaporative flow is established.  The ionized
material flows away, but is constantly replenished by the
photo-evaporation of the disk.  If the star has a strong stellar wind,
the wind may push the confined gas within $r_g$ out to a critical
radius where the ram pressure of the stellar wind is balanced by the
thermal pressure of the photo-evaporated flow, resulting in outflow at
all radii. {\em Yorke and Welz} (1996) and {\em Richling and Yorke}
(1997) made hydrodynamical simulations of the evolution of
photo-evaporated disks under a variety of conditions.  In particular,
{\em Richling and Yorke} found that scattering of ionizing photons on
dust grains increases the photo-evaporation rate.

Recently, {\em Lugo et al.} (2004) modeled  the density and velocity
structure of axisymmetric isothermal winds photo-evaporated from a
spatially thin Keplerian disk.  They calculated the predicted
free-free continuum emission of these models to match the observed
spectral energy distributions of the bipolar objects MWC 349 A and
NGC7538 IRS 1. These models naturally give a bipolar morphology which
is seen in a subset of the HCHII regions. The next step
is to investigate whether the thermally evaporating flow can produce
the high velocities seen in the broad recombination line objects.

\bigskip

\centerline{\textbf{ 4. FUTURE DIRECTIONS}}
\bigskip

A key test of all the proposed models for the HCHII regions and broad
recombination line objects is to match the observed morphologies and
velocity structures of the ionized gas.  Quality datasets already
exist for a few well-studied examples against which to test the
models. At present there are so few objects that
a wider sample is needed to determine whether well-defined morphological
classes actually exist. This may be challenging since they
do appear rare, presumably since the phase is short-lived, but it does
mark an important transition. Further high frequency radio recombination
line observations and, where possible, IR spectroscopy, are needed to
separate those that are merely pressure broadened (and therefore
likely to be understood in terms of the usual UCHII region dynamics)
from those that need a new physical picture.  We also eagerly await the
results currently being obtained with the sub-millimeter array (SMA)
and the future studies that will be made with the millimeter
interferometers CARMA and ALMA. These will be able to trace the
molecular gas dynamics down to the scales of the HCHII
regions.

Once the ionization front expands to UCHII region scales, it is likely
to leave behind the local density enhancement resulting from the
gravitational collapse and begin to experience the wider environment
in which the collapse took place. Significant proper motion relative
to the cloud core will add to this effect.  The presence of champagne
flows in cometary H~II regions indicates that the wider environment must have a
density gradient, i.e. the cometary regions are not located
at the core center.  The original blister picture developed by {\em
Israel} (1978) argued that such regions are a result of triggered star
formation rather than spontaneous collapse; the latter more likely
to occur within a dense core. He was considering more evolved H~II regions,
but they show a similar distribution of morphological types as the
UCHII regions ({\em Fich}, 1993). Velocity studies of large, optically
visible H~II regions have revealed that champagne flows dominate their
dynamics too (e.g., {\em Priestley}, 1999, {\em Roger et al.},
2004). This is most easily understood as the continued evolution of
cometary UCHIIs down a large scale density gradient.

A key test of the dynamical models for cometary regions is to measure
the motion of the stars. Radial velocity measurements on the star in
G29.96-0.02 were attempted by {\em Mart\'{i}n-Hern\'{a}ndez et al.}
(2003). However, the resolution and signal-to-noise of the spectrum,
and uncertainties in the rest wavelengths of the heavy element
spectral features uncontaminated by nebular lines, prevented the
required precision. Once the EVLA and e-MERLIN telescopes come on-line
it may be possible to detect the radio emission directly from the
free-flowing stellar wind of the exciting stars at high
resolution. This would open up the possibility of proper motion
measurements directly on the stars, as well as many more expansion
measurements of the nebulae themselves.

The more global aspects of massive star formation will be addressed by
the numerous galactic plane surveys in coming years, following the
lead by the GLIMPSE survey in the mid-IR. A high resolution radio
survey of the entire northern GLIMPSE region
(www.ast.leeds.ac.uk/Cornish) will pick out all UCHII regions across
the galaxy, building on the previous blind surveys by {\em Giveon et
al.} (2005) and references therein. These will be complemented by a
sub-millimetre continuum survey of the plane at the JCMT, methanol
maser surveys using Parkes and the Lovell telescopes
(www.jb.man.ac.uk/research/methanol), $^{13}$CO from the BU-FCRAO
Galactic Ring Survey (www.bu.edu/GRS), H I from the VLA survey
(www.ras.ucalgary.ca/VGPS) and near-IR from UKIRT (www.ukidss.org).

Unbiased area surveys will allow the luminosity function and lifetimes
to be derived and compared with simulations of the evolution of the
Galactic UCHII region population.  Their location within the wider
GMCs, the fractions that appear to be triggered or isolated and their
location relative to potential external triggers will be established
on a large scale statistical basis. The power of this sensitive,
multi-wavelength, high resolution campaign on the Milky Way will yield
many advances in our understanding of massive star formation.

\textbf{ Acknowledgements.} We would like to thank Jane Arthur for her
input on the hydrodynamic modelling and James Urquhart and Ant
Busfield for help with the figures. Useful discussions were had with
Tom Hartquist and Jim De Buizer. We thank the referee for several
useful suggestions.

\bigskip
%\newpage

\centerline\textbf{ REFERENCES}
\bigskip
\parskip=0pt
{\small
\baselineskip=11pt

\refs Afflerbach A., Churchwell E., Hofner P., and Kurtz S. (1994) {\em Astrophys. J., 437}, 697-704.
\refs Afflerbach, A. et al., (1996) {\em Astrophys. J. Supp., 106}, 423-446.
\refs Alvarez, C. et al., (2004) {\em Astrophys. J. Supp., 155}, 123-148.
\refs Arthur J. and Hoare M. G. (2006) {\em Astrophys. J.}, submitted.
\refs Avalos M., Lizano S., Rodr\'\i guez L., Franco-Hern\'andez R., and Moran J. (2006) {\em Astrophys. J.}, in press.
\refs Benjamin R. A. et al. (2003) {\em Publ. Astron. Soc. Pac., 115}, 953-964.
\refs Bik A., Kaper L., Hanson M. M., and Smits M. (2005) {\em Astron. Astrophys., 440}, 121-137.
\refs Bourke T. L., Hyland A. R., and Robinson G. (2005) {\em Astrophys. J., 625}, 883-890.
\refs Brocklehurst M. and Seaton M. J. (1972) {\em Mon. Not. R. Astron. Soc., 157}, 179-210.
\refs Bunn J. C., Hoare M. G., and Drew J. E. (1995) {\em Mon. Not. R. Astron. Soc.}, 272, 346-354.
\refs Calvet N. et al., (2004) {\em Astron. J., 128}, 1294-1318.
\refs Campbell B., Persson S. E., and McGregor P. J. (1986) {\em Astrophys. J., 305}, 336-352.
\refs Cesaroni R., Churchwell E., Hofner P., Walmsley C. M., and Kurtz S. (1994) {\em Astron. Astrophys., 288}, 903-920.
\refs Cesaroni R., Walmsley C. M, and Churchwell E. (1992) {\em Astron. Astrophys., 256}, 618-630.
\refs Churchwell E. (1999) In {\it The Origin of Stars and Planetary Systems} (C.~Lada and N.~Kylafis, eds.), pp. 515-552. NATO Science Series, Kluwer, The Netherlands.
\refs Churchwell E. (2002) {\em Ann. Rev. Ast. Astrophys.}, 40, 27-62.
\refs Codella C., Felli M., and Natale V. (1994) {\em Astron. Astrophys., 284}, 233-240.
\refs Comeron F. (1997) {\em Astron. Astrophys., 326}, 1195-1214.
\refs Cyganowski C. J., Reid M. J., Fish V. L., and Ho P. T. P. (2003) {\em Astrophys. J. 596}, 344-349.
\refs De Buizer J. M., Radomski J. T., Piña R. K., Telesco C. M. (2002b) {\em Astrophys. J., 580}, 305-316.
\refs De Buizer J. M., Radomski J. T., Telesco C. M., and Piña R. K. (2003) {\em Astrophys. J., 598}, 1127-1139.
\refs De Buizer J. M., Radomski J. T., Telesco C. M., and Piña R. K. (2005) {\em Astrophys. J. Supp, 156}, 179-215.
\refs De Buizer J. M., Watson A. M., Radomski J. T., Piña R. K., and Telesco C. M. (2002a) {\em Astrophys. J., 564}, L101-L104.
\refs De Pree C. G., Rodr\'{i}guez L. F., and Goss W. M.. (1995) {\em Rev. Mex. Astron. Astrophys., 31}, 39-44.
\refs De Pree C. G. et al. (2004) {\em Astrophys. J., 600}, 286-291.
\refs De Pree C. G., Wilner D. J., Deblasio J., Mercer A. J., and Davis L. E. (2005) {\em Astrophys. J., 624}, L101-104.
\refs De Wit W. J., Testi L., Palla F., and Zinnecker H. (2005) {\em Astron. Astrophys., 437}, 247-255.
\refs Dickel H. R. and Goss W. M. (1987) {\em Astron. and Astrophys., 185}, 271-282.
\refs Dickel H. R., Goss W. M., and De Pree C. G. (2001) {\em Astron. J., 121}, 391-398.
\refs Draine B. T. (2003) {\em Ann. Rev. Astron. Astrophys., 41}, 241-289.
\refs Dyson J. and Williams D. (1980) {\it The Physics of the Interstellar Medium}, Wiley, New York.
\refs Dyson J., Williams R., and Redman M. (1995) {\em Mon. Not. R. Astron. Soc., 227}, 700-704.
\refs Ellingsen S. P., Shabala S. S., and Kurtz S. E. (2005) {\em Mon. Not. R. Astron. Soc., 357}, 1003-1012.
\refs Fazio G. G. et al. (1978) {\em Astrophys. J., 221}, L77-81.
\refs Feldt M., Stecklum B., Henning Th., Launhardt R., and Hayward T. L. (1999) {\em Astron. Astrophys., 346}, 243-259.
\refs Felli M., Churchwell E., and Massi M. (1984) {\em Astron. Astrophys., 136}, 53-64.
\refs Fey A. L., Gaume R. A., Claussen M. J., and Vrba F. J. (1995) {\em Astrophys. J., 453}, 308-312.
\refs Fich M. (1993) {\em Astrophys. J. Supp., 86,}, 475-497.
\refs Franco J., et al. (2000) {\em Astrophys. J., 542}, L143-146.
\refs Franco J., Garc\'{i}a-Segura G., and Kurtz S. E  (2006) {\em Astrophys. J.}, submitted.
\refs Franco J., Shore S. N., and Tenorio-Tagle G. (1994) {\em Astrophys. J., 436}, 795-799.
\refs Franco J., Tenorio-Tagle G., and Bodenheimer P. (1990) {\em Astrophys. J., 349}, 126-140.
\refs Franco-Hern\'{n}dez R. and Rodr\'{i}guez L. F. (2004) {\em Astrophys. J., 604}, L105-108.
\refs Fuller G. A. and Myers P. C. (1992) {\em Astrophys. J., 384}, 523-527.
\refs Garay G. and Lizano S. (1999) {\em Publ. Astron. Soc. Pac., 111}, 1049-1087.
\refs Garay G., G\'{o}mez Y., Lizano S., and Brown R.~L. (1998b) {\em Astrophys. J., 501}, 699-709.
\refs Garay G., Lizano S., and G\'{o}mez Y. (1994) {\em Astrophys. J., 429}, 268-284.
\refs Garay G., Lizano S., G\'{o}mez Y., and Brown R.~L. (1998a) {\em Astrophys. J., 501}, 710-722.
\refs Garc\'{i}a-Segura G. and Franco J. (1996) {\em Astrophys. J., 469}, 171-188.
\refs Gaume R., Goss W., Dickel H., Wilson T., and Johnston K. (1995) {\em Astrophys. J., 438}, 776-783.
\refs Giard M., Bernard J. P., Lacombe F., Normand P., and Rouan D. (1994) {\em Astron. Astrophys., 291}, 239-249.
\refs Gies D. R. (1987) {\em Astrophys. J. Supp., 64}, 545-563.
\refs Giveon U., Becker R. H., Helfand D. J., and White R. L. (2005) {\em Astron. J., 129}, 348-354.
\refs G\'{o}mez L., et al. (2005) {\em Astrophys. J., 635}, 1166-1172.
\refs G\'{o}mez Y., et al. (1998) {\em Astrophys. J., 503}, 297-306.
\refs Gonz\'{a}les-Aviles M., Lizano S., and Raga A. C. (2005) {\em Astrophys. J., 621}, 359-371.
\refs Griem H. R. (1974) {\it Spectral Line Broadening by Plasmas},  Academics, New York.
\refs Hatchell J. and van der Tak F. F. S. (2003) {\em Astron. and Astrophys., 409}, 589-598.
\refs Hatchell J., Thompson M. A., Millar T. J., and MacDonald G. H. (1998) {\em Astron. Astrophys.S, 133}, 29-49.
\refs Henning T., Schreyer K., Launhardt R., and Burkert A. (2000) {\em Astron. Astrophys., 353}, 211-226.
\refs Hoare M. G. (2002) In  {\it Hot Star Workshop III: The Earliest Stages of Massive Star Birth}, (P. A. Crowther, ed.),  p137-144. ASP Conf. Ser. 267, San Francisco.
\refs Hoare M. G., et al. (2004) In {\it Milky Way Surveys: The Structure and Evolution of our Galaxy} (D. Clemens, R. Shah, and T. Brainerd, eds.), pp.156-158. Proc. of ASP Conf., 317, San Francisco: ASP.
\refs Hoare M. G., Lumsden S. L., Busfield A. L., Buckley P. 2003 In  {\it Winds, Bubbles and Explosions} (S. J. Arthur and W. J. Henney, eds.),  pp. 172-174. Rev. Mex. Astron. Astrophys. Ser. Conf. 15.
\refs Hoare M. G., Roche P. F., and Glencross W. M. (1991) {\em Mon. Not. R. Astron. Soc., 251}, 584-599.
\refs Hofmann K.-H., Balega Y. Y., Preibisch T., and Weigelt, G. (2004) {\em Astron. Astrophys., 417}, 981-985.
\refs Hofner P. and Churchwell E. (1996) {\em Astron. Astrophys.S, 120}, 283-299.
\refs Hofner P., Delgado H., Whitney B., Churchwell E., and Linz H. (2002) {\em Astrophys. J. 579}, L95-98.
\refs Hofner P., Kurtz S., Churchwell E., Walmsley M., and Cesaroni R. (1996) {\em Astrophys. J. 460}, 359-371.
\refs Hofner P., Kurtz S., Churchwell E., Walmsley C. M., and Cesaroni R. (1994) {\em Astrophys. J., 429}, L85-88.
\refs Hofner P., Wyrowski F., Walmsley C. M., and Churchwell E. (2000) {\em Astrophys. J., 536}, 393-405.
\refs Hollenbach, D., Johnstone, D., Lizano, S., and Shu, F. (1994) {\em Astrophys. J., 428}, 654-669.
\refs Hoogerwerf R., de Bruijne J. H. J., and de Zeeuw P. T. (2001) {\em Astron. Astrophys., 365}, 49-77.
\refs Ignace R. and Churchwell E. (2004) {\em Astrophys. J., 610}, 351-360.
\refs Indebetouw R., et al. (2005) {\em Astrophys. J., 619}, 931-938.
\refs Israel F. P. (1978) {\em Astron. Astrophys., 90}, 769-775.
\refs Jaffe D. and Martin-Pintado J. (1999) {\em Astrophys. J., 520}, 162-172.
\refs Johnson C. O., De Pree C. G., and Goss W. M. (1998) {\em Astrophys. J. 500}, 302-310.
\refs Johnstone D., Hollenbach D., and Bally J. (1998) {\em Astrophys. J., 499}, 758-776.
\refs Kahn F. (1974) {\em Astron. Astrophys., 37}, 149-162.
\refs Kawamura J. H. and Masson C. R. (1998) {\em Astrophys. J., 509}, 270-282.
\refs Keto E. R., Welch W. J., Reid M. J., and Ho P. T. P. (1995) {\em Astrophys. J., 444}, 765-769.
\refs Keto E. (2002a) {\em Astrophys. J., 568}, 754-760.
\refs Keto E. (2002b) {\em Astrophys. J., 580}, 980-986.
\refs Keto E. (2003) {\em Astrophys. J., 599}, 1196-1206
\refs Keto E. and Wood K. (2006) {\em Astrophys. J.}, in press.
\refs Keto E. Welch, W., Reid, M., Ho, P. (1995) {\em Astrophys. J., 444}, 765-769.
\refs Kim K.-T. and Koo B.-C. (2001)  {\em Astrophys. J., 549}, 979-996.
\refs Kim K.-T. and Koo B.-C. (2003)  {\em Astrophys. J., 596}, 362-382.
\refs Kraemer K. E. et al. (2003) {\em Astrophys. J., 588}, 918-930.
\refs Kurtz S., Churchwell E., and Wood D. O. S. (1994) {\em Astrophys. J. Supp., 91}, 659-712.
\refs Kurtz S., et al. (2000) In {\it Protostars and Planets IV} (V. Mannings V., et al., eds.), pp. 299-326. Univ.of Arizona, Tuscon.
\refs Kurtz S. E., Watson A. M., Hofner P., and Otte B. (1999) {\em Astrophys. J., 514}, 232-248.
\refs Lebr\'{o}n M., Rodr\'{i}guez L. F., and Lizano S. (2001) {\em Astrophys. J., 560}, 806-820.
\refs Linz H., Stecklum B., Henning Th., Hofner P., and Brandl B. (2005) {\em Astron. Astrophys., 429}, 903-921.
\refs Lizano S. and Shu F. H. (1989) {\em Astrophys. J., 342}, 834-854.
\refs Lizano S., Cant\'{o} J., Garay G., and Hollenbach D. (1996) {\em Astrophys. J., 468}, 739-748.
\refs Lugo J., Lizano S., and Garay G. (2004) {\em Astrophys. J., 614}, 807-817.
\refs Lumsden S. L. and Hoare M. G. (1996) {\em Astrophys. J., 464}, 272-285.
\refs Lumsden S. L. and Hoare M. G. (1999) {\em Mon. Not. R. Astron. Soc., 305}, 701-706.
\refs Lumsden S. L., Hoare M. G., Oudmaijer R. D., and Richards D. (2002) {\em Mon. Not. R. Astron. Soc., 336}, 621-636.
\refs Lumsden S. L., Puxley P. J., Hoare M. G., Moore T. J. T., and Ridge N. A. (2003) {\em Mon. Not. R. Astron. Soc., 340}, 799-812.
\refs Mac Low M.-M., Van Buren D., Wood D. O. S., and Churchwell  E. (1991) {\em Astrophys. J., 369}, 395-409.
\refs McLaughlin D. E. and Pudritz R. E. (1997) {\em Astrophys. J., 476},  750-765.
\refs Mart\'{i} J., Rodr\'{i}guez L. F., and Reipurth B. (1998) {\em Astrophys. J., 502}, 337-341.
\refs Mart\'{i}n-Hern\'{a}ndez N. L., Bik A., Kaper L., Tielens A. G. G. M., and Hanson M. M. (2003) {\em Astron. Astrophys., 405}, 175-188.
\refs Mart\'{i}n-Hern\'{n}dez N. L., Vermeij R., Tielens A. G. G. M., van der Hulst J. M., and Peeters E. (2002) {\em Astron. Astrophys., 389}, 286-294.
\refs Maxia C., Testi L., Cesaroni R., and Walmsley C. M. (2001) {\em Astron. Astrophys., 371}, 286-299.
\refs Mueller K. E., Shirley Y. L., Evans N. J. II, and Jacobson  H. R. (2002) {\em Astrophys. J. Supp., 143}, 469-497.
\refs Myers P. C. and Fuller G. A. (1992) {\em Astrophys. J., 396}, 631-642.
\refs Natta A. and Panagia N. (1976) {\em Astron. Astrophys., 50}, 191-211.
\refs Nisini B., Smith H. A., Fischer J., and Geballe T. R. (1994) {\em Astron. Astrophys., 290}, 463-472.
\refs Okamoto Y., et al. (2003) {\em Astrophys. J., 584}, 368-384.
\refs Olmi L. and Cesaroni R. (1999) {\em Astron. Astrophys., 352}, 266-276.
\refs Olmi L. et al. (2003) {\em Astron. Astrophys., 407}, 225-235.
\refs Osorio M., Lizano S., and D'Alessio, P. (1999) {\em Astrophys. J., 525}, 808-820.
\refs Patel N. A. et al. (2005) Nature, 437, 109-111.
\refs Pratap P., Megeath S. T., and Bergin E. A. (1999) {\em Astrophys. J., 517}, 799-818.
\refs Priestley C. (1999) PhD Thesis, University of Leeds.
\refs Ramesh B. and Sridharan T. K. (1997) {\em Mon. Not. R. Astron. Soc., 284}, 1001-1006.
\refs Redman M., Williams R., and Dyson J. (1998) {\em Mon. Not. R. Astron. Soc., 298}, 33-41.
\refs Rengarajan T. N. and Ho P. T. P. (1996) {\em Astrophys. J., 465}, 363-370.
\refs Richling S. and Yorke H. W. (1997) {\em Astron. Astrophys., 327}, 317-324.
\refs Rodr\'{i}guez L. F. (1999) in Star Formation 1999, Ed. T. Nakamoto, Nobeyama Radio Observatory, p. 257-262.
\refs Roger R. S., McCutcheon W. H., Purton C. R., and Dewdney P. E. (2004) {\em Astron. Astrophys., 425}, 553-567.
\refs Roshi A., Balser D. S., Bania T. M., Goss W. M., and De Pree C. G. (2005) {\em Astrophys. J., 625}, 181-193.
\refs Sewilo M., Churchwell E., Kurtz S., Goss W., and Hofner P. (2004) {\em Astrophys. J. 609}, 285-299.
\refs Sewilo M., et al. (2005) In {\it Massive Star Birth: A Crossroads in Astrophysics}, IAU Symp. 227 (R. Cesaroni et al., eds.) Poster Proceedings \\ www.arcetri.astro.it/iaus227/posters/sewilo\_m.pdf.
\refs Shu F. (1992) {\it The Physics of Astrophysics, Volume II, Gas Dynamics}, University Science Books, Mill Valley, CA.
\refs Shull J. M. (1980) {\em Astrophys. J., 238}, 860-866.
\refs Sicilia-Aguilar A. et al. (2005) {\em Astron. J., 129}, 363-381.
\refs Sim S. A., Drew J. E., and Long K. S. (2005) {\em Mon. Not. R. Astron. Soc., 363}, 615-627.
\refs Smith L. J., Norris R. P. F., and Crowther P. A. (2002)  {\em Mon. Not. R. Astron. Soc., 337}, 1309-1328.
\refs Smith N. et al. (2000) {\em Astrophys. J., 540}, 316-331.
\refs Snell R. L. and Bally J. (1986) {\em Astrophys. J., 303}, 683-701. 
\refs Sollins P., Zhang Q., Keto E., and Ho P. (2004) {\em Astrophys. J., 624}, L49-52.
\refs Sollins P., Zhang Q., Keto E., and Ho P. (2005) {\em Astrophys. J., 631}, 399-410.
\refs Spitzer L., Jr. (1978) {\it Physical Processes in the Interstellar Medium}, Wiley, New York.
\refs Stecklum B. et al. (2002) {\em Astron. Astrophys., 392}, 1025-1029.
\refs Tan J. C. and McKee C. F. (2003) In {\it Star Formation at High Angular Resolution}, IAU Symp. 221, (M. G. Burton et al., eds.), ASP, astro-ph/0309139).
\refs Terebey S., Shu F., and Cassen P., (1984) {\em Astrophys. J., 286}, 529-551.
\refs Turner B. E. and Matthews H. E. (1984) {\em Astrophys. J., 277}, 164-180.
\refs Ulrich R. (1976) {\em Astrophys. J., 210}, 377-391.
\refs van der Walt D. J. (1997) {\em Astron. Astrophys., 322}, 307-310.
\refs Walsh A. J., Burton M. G., Hyland A. R., and Robinson G. (1998) {\em Mon. Not. R. Astron. Soc., 301}, 640-698.
\refs Watson A. M. and Hanson M. M. (1997) {\em Astrophys. J., 490}, L165-169.
\refs Watt S. and Mundy L. G. (1999) {\em Astrophys. J. Supp., 125}, 143.-160
\refs Williams R. J. R., Dyson J. E., and Hartquist T. W. (2000) 314, 315-323.
\refs Wolfire M. and Cassinelli. J. (1987) {\em Astrophys. J., 319}, 850-867.
\refs Wood D. O. S. and Churchwell E. (1989a) {\em Astrophys. J. Supp., 69}, 831-895.
\refs Wood D. O. S. and Churchwell E. (1989b) {\em Astrophys. J., 340}, 265-272.
\refs Wood D. O. S. and Churchwell E. (1991) {\em Astrophys. J., 372}, 199-207.
\refs Wright A. E. and Barlow M. J. (1975) {\em Mon. Not. R. Astron. Soc., 170}, 41-51.
\refs Wyrowski F., Schilke P., Walmsley C. M., and Menten K. M. (1999) {\em Astrophys. J., 514}, L43-46.
\refs Xie T., Mundy L. G., Vogel S. N., and Hofner P. (1996) 473, L131-134.
\refs Yorke H. W. (1986) {Ann. Rev. Astron. Astrophys., 24}, 49-87.
\refs Yorke H. W. and Welz A. (1996) {\em Astron. Astrophys., 315}, 555-564.
\refs Yorke H. W., Tenorio-Tagle G., Bodenheimer P., 1983, {\em Astron. Astrophys., 127}, 313-319.
\refs Zhu Q.-F., Lacy J. H., Jaffe D. T., Greathouse T. K., and Richter M. J. (2005) {\em Astrophys. J., 631}, 381-398.

\end{document}